\documentclass[showpacs,aps,prd,nofootinbib,floatfix,amsmath,amssymb]{revtex4}
\usepackage{graphicx}
\begin{document}

\makeatletter
\newbox\slashbox \setbox\slashbox=\hbox{$/$}
\newbox\Slashbox \setbox\Slashbox=\hbox{\large$/$}
\def\pFMslash#1{\setbox\@tempboxa=\hbox{$#1$}
  \@tempdima=0.5\wd\slashbox \advance\@tempdima 0.5\wd\@tempboxa
  \copy\slashbox \kern-\@tempdima \box\@tempboxa}
\def\pFMSlash#1{\setbox\@tempboxa=\hbox{$#1$}
  \@tempdima=0.5\wd\Slashbox \advance\@tempdima 0.5\wd\@tempboxa
  \copy\Slashbox \kern-\@tempdima \box\@tempboxa}
\def\FMslash{\protect\pFMslash}
\def\FMSlash{\protect\pFMSlash}
\def\miss#1{\ifmmode{/\mkern-11mu #1}\else{${/\mkern-11mu #1}$}\fi}
\makeatother

\title{Gauge invariance and quantization of Yang--Mills theories in extra dimensions}
\author{H. Novales--S\' anchez and J. J. Toscano}
\address{Facultad de Ciencias F\'{\i}sico Matem\'aticas,
Benem\'erita Universidad Aut\'onoma de Puebla, Apartado Postal
1152, Puebla, Puebla, M\'exico.}
\begin{abstract}
The gauge structure of the four dimensional effective theory arising from a pure $SU_5(N)$ Yang-Mills theory in five dimensions compactified on the orbifold $S^1/Z_2$ is reexamined on the basis of the Becchi--Rouet--Stora--Tyutin (BRST) symmetry. In this context, the two scenarios that can arise are analyzed: if the gauge parameters propagate in the bulk, the excited Kaluza--Klein (KK) modes are gauge fields, but they are matter vector fields if these parameters are confined in the 3--brane. In the former case, it is shown that the four dimensional theory is gauge invariant only if the compactification is carried out by using curvatures instead of gauge fields as fundamental objects. Then, it is shown that the four dimensional theory is governed by two types of gauge transformations, one determined by the KK zero modes of the gauge parameters, $\alpha^{(0)a}$, and the other by the excited KK modes, $\alpha^{(n)a}$. The Dirac's method and the proper solution of the master equation in the context of the field-antifield formalism are employed to show that the theory is subject to first class constraints. A gauge-fixing procedure to quantize the KK modes $A^{(n)a}_\mu$ that is covariant under the first type of gauge transformations, which embody the standard gauge transformations of $SU_4(N)$, is introduced through gauge-fixing functions transforming in the adjoint representation of this group. The ghost sector induced by these gauge-fixing functions is derived on the basis of the BRST formalism. The effective quantum Lagrangian that links the interactions between light physics (zero modes) and heavy physics (excited KK gauge modes) is presented. Concerning the radiative corrections of the excited KK modes on the light Green functions, the predictive character of this Lagrangian at the one--loop level is stressed. In the case of the gauge parameters confined to the 3-brane, the known result in the literature is reproduced with some minor variants, although it is emphasized that the exited KK modes are not gauge fields, but matter fields that transform under the adjoint representation of $SU_4(N)$. The Dirac's method is employed to show that this theory is subject to both first and second class constraints, which arise from the zero and excited KK modes, respectively.
\end{abstract}


\maketitle

\section{Introduction}
\label{Int}Theories which involve more than four dimensions received renewed interest due to developments in supergravity and superstring theories three decades ago. However, the extra dimensions contemplated then are extremely small, of the order of the inverse Planck scale $M^{-1}_{Pl}$, to be of some phenomenological interest. It was only after the pioneering works by Antoniadis, Arkani--Hamed, Dimopoulos and Dvali~\cite{antoniadis,AHDD,AADD}, where large extra dimensions were considered, that extra dimensions became phenomenologically attractive. In most scenarios, our observed 3--dimensional space is a 3--brane and is embedded in a higher $D$--dimensional spacetime, which is known as the bulk. If the additional dimensions are small enough, the Standard Model (SM) gauge and matter fields are phenomenologically allowed to propagate in the bulk; otherwise they are stuck to the 3--brane. Of course, if there are extra dimensions, they must be smaller than the smallest scale which has been currently explored by experiments. So, the extra dimensions are assumed to be suitably compactified on some manifold whose size is sufficiently small. As a result of the compactification, those fields that propagate in the bulk expand into a series of states known as a KK tower, with the individual KK excitations being labeled by mode numbers.

In the last decade, the phenomenological implications of extra dimensions on low--energy observables have been the subject of considerable interest. However, to our knowledge, the internal consistence of the gauge sector of the four dimensional effective theory still remains unclear. Although some authors~\cite{Dienesetal,W1,W2,Chivukulaetal,chinos,W3,W4,W5} have worked out Yang--Mills theories in five dimensions ($SU_5(N)$), it is not clear how the four dimensional theory is governed by the $SU_4(N)$ gauge group, since there is an infinite number of gauge parameters, namely the tower of KK modes, $\alpha^{(n)a}(x)$, which arise from assuming that the parameters of $SU_5(N)$, $\alpha^a(x,y)$, propagate in the bulk. Although the gauge transformations to which the excited $A^{(n)a}_\mu(x)$ KK modes are subject have been in part derived~\cite{W5}, it is doubtful that the Lagrangian there obtained respects such a set of gauge transformations. The only gauge symmetry that becomes manifest is the standard one, which is characterized by the zero modes, $\alpha^{(0)a}(x)$, of the five dimensional gauge parameters. In the context of these \textit{standard gauge transformations} (SGT), the covariant objects are easily identified, as they arise naturally once the compactification and integration of the extra dimension is carried out. The zero mode, $A^{(0)n}_\mu(x)$, of the five dimensional gauge field ${\cal A}^a_M(x,y)$ corresponds to the standard Yang--Mills field, whereas the covariant objects are the usual curvature $F^{(0)a}_{\mu \nu}(x)$, the excited KK modes $A^{(n)a}_\mu(x)$, and the pseudo Goldstone bosons $A^{(n)a}_5(x)$, which transform in the adjoint representation of $SU_4(N)$. However, no similar objects have been identified for the \textit{nonstandard gauge transformations} (NSGT) characterized by the  $\alpha^{(n)a}(x)$ parameters and does not seem easy to construct them by direct observation of the four dimensional Lagrangian, which even differs from one to other author. In this work, we will show that such four dimensional Yang--Mills theories are not  invariant indeed under the NSGT but only under the SGT. Below, we will show that this type of four dimensional theories can arise from a five dimensional theory in which is assumed that the gauge parameters are confined to the 3--brane. In such a scenario, the excited KK modes $A^{(n)a}_\mu(x)$ are not gauge fields, but massive vector fields, whereas the scalar fields $A^{(n)a}_5(x)$ do not correspond to pseudo Goldstone bosons, but they are massless scalar fields. When the gauge parameters are allowed propagate in the bulk, the excitations $A^{(n)a}_\mu(x)$ and $A^{(n)a}_5(x)$ correspond to gauge fields and pseudo Goldstone bosons, respectively. However, the four dimensional theory is described by a Lagrangian which differs substantially from those given in the literature. In this work, we will derive a four dimensional Lagrangian which is separately invariant under both the SGT and the NSGT. This Lagrangian will be written out in terms of covariant objets that resemble the standard Yang--Mills curvature. The discrepancy between our result and the ones given in the literature comes from the way that the integration on the fifth dimension is carried out in the action representing the five dimensional theory. We will show that for the scenario with gauge parameters propagating in the bulk, gauge invariance under the NSGT is lost unless the integration of the fifth dimension contemplates only the Fourier modes of five--dimensional covariant objets. This means that the curvatures are the objects which must be expanded in Fourier series and not the gauge fields, which is the method most commonly followed in the literature so far. On the contrary, if the gauge parameters do not propagate in the fifth dimension, one can carry out the Fourier expansion at the level of fields, since in this case the NSGT does not arise. Although this scenario is less interesting from the physical point of view, as it contains massless scalar fields, we will also consider it for comparison purposes. Of course, we will center our discussion in the scenario in which the gauge parameters propagate in the bulk, as it allows us to make contact with the physical reality. In particular, the identification of NSGT that arise in this scenario and the construction of a classical action being invariant under it, is an important goal of this work.

As commented above, so far there is not a consistent four dimensional KK Yang--Mills theory that includes a precise description of the gauge symmetries to which it is subject. However, the precise identification of these gauge transformations, as well as the covariant objets needed to construct invariants, is a first indispensable step to quantize the theory. While most of the studies have been restricted to tree--level processes, the quantum loop effects of the theory have received much less attention, as only some one--loop processes, as electromagnetic dipoles~\cite{ED}, the $b\to s\gamma$~\cite{BSG}, $Z\to \bar{b}b$~\cite{Papa,ZBB}, $B_{s,d}\to \gamma \gamma$~\cite{BsdGG}, $B_d\to l^{+}l^{-}$~\cite{Bdll} decays and $B^{0}-\bar{B}^{0}$~\cite{B0B0} mixing have been considered. Nonetheless, to our knowledge, no quantum fluctuations with only gauge KK modes circulating in the loops have been considered so far. To predict this class of effects one needs of a theory that is consistently quantized. As it is well known, to quantize the theory a gauge--fixing procedure for the gauge fields must be introduced and their associated ghost sector derived. In order to do this, the precise gauge transformations obeyed by the gauge degrees of freedom, in our case the KK modes associated with the Yang--Mills fields, must be known in a precise way. The modern approach to quantize gauge systems based in the BRST symmetry~\cite{BRST} requires the ghost fields, which coincide with the parameters of the gauge group but with opposite statistics, to be introduced from the beginning~\cite{PR}. This procedure is well known in standard Yang--Mills theories, but its implementation for the case of the NSGT is more elaborated. The first step consists in determining the tensorial structure of the NSGT. At least there are three equivalent ways of knowing them. One way consists in deriving them directly from the infinitesimal gauge transformation obeyed by the gauge fields in five dimensions. To carry out this it is only necessary to specify the periodicity and the parity of the gauge fields and the gauge parameters with respect to the compactified coordinate. Another different method to find the NSGT consists in deriving the constraints of the theory employing the Dirac's method and then use the Castellani's gauge generator~\cite{Castellani}, which allows us to determine the gauge transformations of the gauge fields. The third way, which is intimately related to the latter, consists in using the field--antifield formalism~\cite{PR}. In this formulation, the BRST symmetry arises naturally through the introduction of the antibracket~\cite{PR}. To be sure of the precise gauge structure of the four dimensional KK Yang--Mills theory, we will employ all these schemes. In particular, we will use the proper solution of the master equation in five dimensions to derive the corresponding proper solution in four dimensions. Then, we will use this solution to define the gauge--fixed action with respect to the NSGT, $S_{\Psi^{NSGT}}$, through a gauge--fixing fermion functional,$\Psi^{NSGT}$, that is invariant under the SGT of $SU_4(N)$.

There are phenomenological and theoretical motivations to quantize a gauge KK theory. If KK modes cannot be produced directly in the LHC collider, it would be possible to detect their virtual effects through precision measurements as those planed to be realized in the International Linear Collider~\cite{LHC-ILC}. Electroweak precision observables can play a role in various models. In many physics scenarios they can provide information about new physics scales that are too heavy to be detected directly. Due to this, it is important to count on a consistent quantum theory of the KK excitations that allows us to perform predictions at the one--loop or higher orders. In particular, it is important to calculate the one--loop effects of these new particles on SM observables that eventually could be sensitive to new physics effects. On the theoretical side, it is interesting to investigate the behavior of the theory at the one--loop level. For instance, it is very important to study the UV structure of light Green functions, \textit{i.e.} Green functions consisting of zero modes only, due to loop contributions of excited modes. This is an important objective of this work. As already commented, the physical $A^{(n)a}_\mu(x)$ fields are subject to satisfy complicated NSGT. So, to quantize these gauge fields one necessarily must invoke a gauge--fixing procedure. The fact that these fields also obey the SGT greatly facilitate the things, as it allows us to introduce a gauge--fixing procedure that respects the SGT and thus to have an highly symmetric effective quantum action. We will show below that it is possible to define the propagators of these particles by introducing nonlinear gauge--fixing functions transforming covariantly under the SGT of the $SU_4(N)$ gauge group. One important goal of this work is to show that the one--loop effects of the KK modes on light Green functions are perfectly calculable like in any renormalizable theory beyond the SM. Let us clarify this point. As already mentioned, once the fifth dimension is compactified and integrated, the theory can be written in terms of an effective Lagrangian, which can be arranged as follows:
\begin{equation}
\label{eq1}
{\cal L}_{ED}={\cal L}^{(0)}\left(A^{(0)a}_\mu\right)+{\cal L}^{(0)(n)}\left(A^{(0)a}_\mu,A^{(n)a}_\mu,A^{(n)a}_5\right)+{\cal L}^{(n)}\left(A^{(n)a}_\mu,A^{(n)a}_5\right)\, ,
\end{equation}
where the first term ${\cal L}^{(0)}$ represents the usual renormalizable Yang--Mills theory, the second term ${\cal L}^{(0)(n)}$ contains the interactions of the usual Yang--Mills fields with the KK modes. Finally, the third term ${\cal L}^{(n)}$ involves only interactions between excited KK modes. Our point is that the one--loop contribution to light Green functions is perfectly determined like in any renormalizable theory. As we will discuss below, the structure of the quantized theory suggests that the only divergences induced by the excited KK modes at the one--loop level on light Green functions are those already present in the standard Yang--Mills theory and can therefore be absorbed by the parameters of the light theory. We will show this explicitly by using the Background Field Method~\cite{BFM} to quantize the standard Yang--Mills theory, as in this scheme gauge invariance with respect to the SGT is preserved. As it is well known, this sort of gauge invariance sets powerful constraints on the infinities that can occur in the $\Gamma$ effective action, being particularly simple at the one--loop level. We will see that, as a consequence of our gauge--fixing procedure for the KK modes, which is covariant under the SGT, the type of infinities generated by the KK modes at the one--loop level are identical to those generated by the zero modes when quantized using the background field gauge, which then allows us to absorb them in the parameters of the light theory. It is important to point out that this well behavior of light Green functions at the one--loop level does not mean that they are also renormalizable at the two--loop level or higher orders, and by no means that the complete theory is renormalizable, as it is well known that gauge theories in more than four dimensions are not renormalizable in the Dyson's sense. So they must be recognized as effective theories that become embedded in some other consistent UV completion, such as string theories. However, it should be mentioned that effective field theories are renormalizable in a wider sense~\cite{WBooks,Examples}. Below, we will present some comments concerning the possible variants that will arise in the context of an effective theory of this type. The nonrenormalizable nature of higher dimensional theories arises from the fact that they have dimensionfull coupling constants. So, the effective theory must be cut off at some scale $M_s$, above which the fundamental theory enters. The cutoff insensitivity of light Green functions at the one--loop level, which seems to be exclusive of the so-called universal extra dimensional (UED) models (theories in which all the fields propagate in the extra dimensions) with only one extra dimension, has been already pointed out in previous studies on some electroweak observables~\cite{Papa,Apel}. In this work, we will show that this is the case for a pure Yang--Mills theory in five dimensions.

The rest of the paper has been organized as follows. In Sec. \ref{Com}, the  five--dimensional $SU_5(N)$ theory is discussed. A compactification scheme is defined and the gauge structure of the four--dimensional theory determined. Both the standard and nonstandard gauge transformations are determined and covariant objects under these gauge transformations identified. Sec. \ref{Q} is devoted to quantize the theory. The proper solution of the master equation in five dimensions is used to derive the corresponding proper solution for the four dimensional theory. A gauge--fixing procedure for the excited KK modes that is covariant under standard gauge transformations is introduced. Then, the nonstandard gauge transformations are used to determine the ghost Lagrangian. In Sec. \ref{olr}, the renormalizability of the light Green functions at the one--loop level is discussed. Finally, in Sec. \ref{C} the conclusions are presented.

\section{A pure $SU(N)$ theory in five dimensions}
\label{Com} This section is devoted to study the gauge structure of the four dimensional theory that arises after carrying out the compactification of a five dimensional pure $SU_5(N)$ theory. The compactification conditions will be defined and the corresponding four dimensional Lagrangian derived. The main result of this section will be the derivation of the four dimensional Lagrangian together with the SGT and the NSGT, already commented in the introduction, to which it is subject.

\subsection{Compactification}
\label{L4YM}
In the following, we will denote by $x$ the usual four coordinates and by $y$ the one that corresponds to the fifth dimension. We will employ a flat metric with signature $diag(1,-1,-1,-1,-1)$. The gauge fields will be denoted by $A^a_M(x,y)$, with $M(=0,1,2,3,5)$ and $a$ the Lorentz and gauge indices, respectively. As usual, Greek indices will be used to denote the four dimensional spacetime. Consider the five dimensional Yang--Mills theory given by the following action
 \begin{equation}
 \label{S0}
 S_0=\int d^4x \int dy \, {\cal L}_{5YM}(x,y)\, ,
 \end{equation}
 where the five dimensional Lagrangian is given by
 \begin{equation}
 {\cal L}_{5YM}(x,y)=-\frac{1}{4}{\cal F}^a_{MN}(x,y){\cal F}^{MN}_a(x,y)\, ,
 \end{equation}
 with the curvature defined in terms of the gauge fields ${\cal A}^a_M$ as follows
 \begin{equation}
{\cal F}^a_{MN}=\partial_M{\cal A}^a_N-\partial_N {\cal A}^a_M+g_5f^{abc}{\cal A}^b_M{\cal A}^c_N\, .
 \end{equation}
 Notice that the coupling constant $g_5$ has dimension of $(mass)^{-1/2}$. As commented in the introduction, two scenarios arise depending on whether the gauge parameters propagate in the bulk or not. We first consider the case of gauge parameters propagating in the fifth dimension. The gauge fields transform as
 \begin{equation}
 \label{TLA}
 \delta {\cal A}^a_{M}={\cal D}^{ab}_M \alpha^b(x,y),
 \end{equation}
 where ${\cal D}^{ab}_M=\delta^{ab}\partial_M-g_5f^{abc}{\cal A}^c_M$ is the covariant derivative in the adjoint representation of $SU_5(N)$ and $\alpha^a(x,y)$ are the gauge parameters. The covariant object in this theory is the curvature, which transforms in the adjoint representation:
 \begin{equation}
 \label{TLF}
 \delta {\cal F}^a_{MN}=g_5f^{abc}{\cal F}^b_{MN}\alpha^c(x,y)\,.
 \end{equation}
It should be noticed that up to this point, all the spatial dimensions have been treated equally.

We now assume that the fifth dimension is compactified on a $S^1/Z_2$ orbifold whose radius is denoted by $R$. This choice imposes some periodic and parity conditions on the gauge fields and gauge parameters with respect to the extra dimension. As it was emphasized in the introduction, the gauge structure of the four dimensional theory depends crucially on how the Fourier expansions are performed in the integral
\begin{eqnarray}
\label{L4}
{\cal L}_{4YM}&=&-\frac{1}{4}\int^{2\pi R}_0 {\cal F}^a_{MN}(x,y){\cal F}^{MN}_a(x,y)\nonumber \\
&=&-\frac{1}{4}\int^{2\pi R}_0 \left[{\cal F}^a_{\mu \nu}(x,y){\cal F}^{\mu \nu}_a(x,y)+2{\cal F}^a_{\mu 5}(x,y){\cal F}^{\mu 5}_a(x,y)\right]\, ,
\end{eqnarray}
where
\begin{eqnarray}
\label{def1}
{\cal F}^a_{\mu \nu}(x,y)&=&\partial_\mu{\cal A}^a_\nu-\partial_\nu {\cal A}^a_\mu+g_5f^{abc}{\cal A}^b_\mu{\cal A}^c_\nu \, ,\\
\label{def2}
{\cal F}^a_{\mu 5}(x,y)&=&\partial_\mu{\cal A}^a_5-\partial_5 {\cal A}^a_\mu+g_5f^{abc}{\cal A}^b_\mu{\cal A}^c_5\, .
\end{eqnarray}
As it will clear later on, gauge invariance only is preserved after compactification and integration of the fifth dimension if the Fourier expansions are implemented at the level of the curvatures ${\cal F}^a_{\mu \nu}(x,y)$ and ${\cal F}^a_{\mu 5}(x,y)$ and not at the level of the fields ${\cal A}^a_\mu$ and ${\cal A}^a_5$, as it has been made in the literature. Accordingly, we implement the following periodic and parity conditions:
\begin{eqnarray}
{\cal F}^a_{MN}(x,y+2\pi R)&=&{\cal F}^a_{MN}(x,y)\, ,\\
\alpha^a(x,y+2\pi R)&=&\alpha^a(x,y)\, ,
\end{eqnarray}
\begin{eqnarray}
{\cal F}^a_{\mu \nu}(x,-y)&=&{\cal F}^a_{\mu \nu}(x,y)\, ,\\
{\cal F}^a_{\mu 5}(x,-y)&=&-{\cal F}^a_{\mu 5}(x,y)\, ,\\
\alpha^a(x,-y)&=&\alpha^a(x,y)\, .
\end{eqnarray}
These periodicity and parity properties of the curvatures and gauge parameters allow us to expand them in Fourier series as follows:
\begin{eqnarray}
\label{ec1}
{\cal F}^a_{\mu \nu}(x,y)&=&\frac{1}{\sqrt{2\pi R}}{\cal F}^{(0)a}_{\mu \nu}(x)+\sum_{m=1}^\infty \frac{1}{\sqrt{\pi R}}{\cal F}^{(m)a}_{\mu \nu}(x)\cos\left(\frac{my}{R}\right)\, ,\\
\label{ec2}
{\cal F}^a_{\mu 5}(x,y)&=&\sum_{m=1}^\infty \frac{1}{\sqrt{\pi R}}{\cal F}^{(m)a}_{\mu 5}(x)\sin\left(\frac{my}{R}\right)\, ,
\end{eqnarray}
\begin{equation}
\alpha^a(x,y)=\frac{1}{\sqrt{2\pi R}}\alpha^{(0)a}(x)+\sum_{m=1}^\infty \frac{1}{\sqrt{\pi R}}\alpha^{(m)a}(x)\cos\left(\frac{my}{R}\right)\, .
\end{equation}
It is important to stress at this point the fact that although the gauge parameters do not participate dynamically at the classical level, they play a crucial role at the quantum level. They become the Faddeev--Popov ghost fields, but with opposite statistics. Indeed, the modern approach to the BRST symmetry~\cite{PR} promotes these fields as true degrees of freedom from the beginning, at the same level of the gauge fields, as they are needed to quantize the theory. The importance of investigating the role played by the gauge parameters in the four-dimensional theory, to proceed then to quantize it, is now clear.

Once replaced the curvatures ${\cal F}^a_{\mu \nu}(x,y)$ and ${\cal F}^a_{\mu 5}(x,y)$ into the integral (\ref{L4}) by their respective Fourier series, one obtains
\begin{equation}
\label{lag}
{\cal L}_{4YM}=-\frac{1}{4}\left({\cal F}^{(0)a}_{\mu \nu}{\cal F}^{(0)a\mu \nu}+{\cal F}^{(m)a}_{\mu \nu}{\cal F}^{(m)a\mu \nu}+2\, {\cal F}^{(m)a}_{\mu 5}{\cal F}^{(m)a\mu 5}\right)\, ,
\end{equation}
where sums over all repeated indices, including the Fourier ones, are assumed. From now on, this convention will be maintained through the paper. Notice that due to orthogonality of the trigonometric functions there is no interference between zero modes and excited modes of the curvatures.

In order to determine explicitly the form of the curvatures ${\cal F}^{(0)a}_{\mu \nu}$, ${\cal F}^{(m)a}_{\mu \nu}$, and ${\cal F}^{(m)a}_{\mu 5}$, we use the definitions of the tensors ${\cal F}^a_{\mu \nu}(x,y)$ and ${\cal F}^a_{\mu 5}(x,y)$, given by Eqs.(\ref{def1},\ref{def2}), and expand in Fourier series the gauge fields that constitute them, which leads to
\begin{eqnarray}
\label{ec3}
{\cal F}^a_{\mu \nu}(x,y)&=&\frac{1}{\sqrt{2\pi R}}F^{(0)a}_{\mu \nu}(x)+\sum_{m=1}^\infty \frac{1}{\sqrt{\pi R}}\left({\cal D}^{(0)ab}_\mu A^{(m)b}_\nu-{\cal D}^{(0)ab}_\nu A^{(m)a}_\mu\right)\cos\left(\frac{my}{R}\right)\nonumber \\
&&+gf^{abc}\sum_{m=1}^\infty \sum_{n=1}^\infty \sqrt{\frac{2}{\pi R}}A^{(m)b}_\mu A^{(n)c}_\nu \cos\left(\frac{my}{R}\right)\cos\left(\frac{ny}{R}\right)\, ,
\end{eqnarray}
\begin{eqnarray}
\label{ec4}
{\cal F}^a_{\mu 5}(x,y)&=&\sum_{m=1}^\infty \frac{1}{\sqrt{\pi R}}\left({\cal D}^{(0)ab}_\mu A^{(m)b}_5+\frac{m}{R}A^{(m)a}_\mu \right)\sin\left(\frac{my}{R}\right)\nonumber \\
&&+gf^{abc}\sum_{m=1}^\infty \sum_{n=1}^\infty \sqrt{\frac{2}{\pi R}}A^{(m)b}_\mu A^{(n)c}_5 \cos\left(\frac{my}{R}\right)\sin\left(\frac{ny}{R}\right)\, ,
\end{eqnarray}
where
\begin{equation}
F^{(0)a}_{\mu \nu}=\partial_\mu A^{(0)a}_\nu-\partial_\nu A^{(0)a}_\mu+gf^{abc}A^{(0)b}_\mu A^{(0)c}_\nu \, .
\end{equation}
In addition, ${\cal D}^{(0)ab}_\mu=\delta^{ab}\partial_\mu-gf^{abc}A^{(0)c}_\mu$ and $g=g_5/\sqrt{2\pi R}$. As a next step, we equalize the right--hand side members of Eqs. (\ref{ec1}) and (\ref{ec3}). Then, we multiply by $1/\sqrt{2\pi R}$, integrate over $y$ in the interval $0\leq y\leq 2\pi R$ and use the orthogonality of the trigonometric functions to obtain
\begin{equation}
{\cal F}^{(0)a}_{\mu \nu}=F^{(0)a}_{\mu \nu}+gf^{abc}A^{(m)b}_\mu A^{(m)c}_\nu \, .
\end{equation}
Multiplying now by $(1/\sqrt{\pi R})\cos\left(\frac{ny}{R}\right)$ and proceeding in the same way leads to
\begin{equation}
{\cal F}^{(m)a}_{\mu \nu}={\cal D}^{(0)ab}_\mu A^{(m)b}_\nu-{\cal D}^{(0)ab}_\nu A^{(m)b}_\mu +gf^{abc}\Delta^{mrn}A^{(r)b}_\mu A^{(n)c}_\nu \,
\end{equation}
where
\begin{equation}
\Delta^{mrn}=\frac{1}{\sqrt{2}}\left(\delta^{r,m+n} +\delta^{m,r+n}+\delta^{n,r+m}\right)\, .
\end{equation}
Finally, we equalize the right--hand side members of Eqs. (\ref{ec2}) and (\ref{ec4}) and follow the same path as in the previous case to obtain
\begin{equation}
{\cal F}^{(m)a}_{\mu 5}={\cal D}^{(0)ab}_\mu A^{(m)b}_5+\frac{m}{R}A^{(m)a}_\mu +gf^{abc}\Delta'^{mrn}A^{(r)b}_\mu A^{(n)c}_5 \, ,
\end{equation}
where
\begin{equation}
\Delta'^{mrn}=\frac{1}{\sqrt{2}}\left(\delta^{m,r+n} +\delta^{r,m+n}-\delta^{n,r+m}\right) \, .
\end{equation}

The gauge variation of the four dimensional ${\cal F}^{(0)a}_{\mu \nu}$, ${\cal F}^{(m)a}_{\mu \nu}$, and ${\cal F}^{(m)a}_{\mu 5}$ curvatures is encoded in the corresponding gauge variation for the five dimensional curvature ${\cal F}^a_{MN}$ given by Eq.(\ref{TLF}), which can be decomposed into two parts as follows
\begin{eqnarray}
\label{TLF1}
\delta {\cal F}^a_{\mu \nu}(x,y)&=&g_5f^{abc}{\cal F}^{b}_{\mu \nu}(x,y)\alpha^c(x,y)\, ,\\
\label{TLF2}
\delta {\cal F}^a_{\mu 5}(x,y)&=&g_5f^{abc}{\cal F}^{b}_{\mu 5}(x,y)\alpha^c(x,y)\, .
\end{eqnarray}
Expanding both the left--hand and the right--hand sides of these equations in Fourier series and using the orthogonality of the trigonometric functions, one obtains
\begin{equation}
\label{lt1}
\delta {\cal F}^{(0)a}_{\mu \nu}=gf^{abc}\left({\cal F}^{(0)b}_{\mu \nu}\alpha^{(0)c}+{\cal F}^{(m)b}_{\mu \nu}\alpha^{(m)c}\right)\, ,
\end{equation}
\begin{equation}
\label{lt2}
\delta {\cal F}^{(m)a}_{\mu \nu}=gf^{abc}\left({\cal F}^{(m)b}_{\mu \nu}\alpha^{(0)c}+\left(\delta^{mn}{\cal F}^{(0)b}_{\mu \nu}+\Delta^{mrn}{\cal F}^{(r)b}_{\mu \nu} \right)\alpha^{(n)c}\right)\, ,
\end{equation}
\begin{equation}
\label{lt3}
\delta {\cal F}^{(m)a}_{\mu 5}=gf^{abc}\left({\cal F}^{(m)b}_{\mu 5}\alpha^{(0)c}+\Delta'^{mrn}{\cal F}^{(r)b}_{\mu 5}\alpha^{(n)c}\right) \, .
\end{equation}
It is not difficult to show that the ${\cal L}_{4YM}$ Lagrangian is invariant under these transformations of the curvatures. However, it is important to stress that this Lagrangian differs from those presented in the literature~\cite{W1,W2,Chivukulaetal,W3,W4,W5}.

On the other hand, it is interesting to investigate the structure of the equations of motion, which also serves to check on the ${\cal L}_{4YM}$ Lagrangian, as the equations of motion can be derived in essentially two different ways, namely, directly from the four dimensional theory characterized by this Lagrangian and by compactification of the five dimensional equations of motion. The equations of motion coming from the four dimensional theory are given by
\begin{equation}
\partial_\mu \left(\frac{\partial {\cal L}_{4YM}}{\partial_\mu A}\right)=\frac{\partial {\cal L}_{4YM}}{\partial A}\, ,
\end{equation}
where $A$ stand for $A^{(0)a}_\nu$, $A^{(m)a}_\nu$, and $A^{(m)a}_5$. A straightforward calculation leads to
\begin{equation}
{\cal D}^{(0)ab}_\mu {\cal F}^{(0)b\mu \nu}=gf^{abc}\left({\cal F}^{(m)b\nu 5}A^{(m)c}_5+{\cal F}^{(m)b\mu \nu}A^{(m)c}_\mu \right) \, ,
\end{equation}
\begin{equation}
{\cal D}^{(mn)ab}_\mu {\cal F}^{(n)b\mu \nu}=gf^{abc}{\cal F}^{(0)b\mu \nu}A^{(m)c}_\mu-{\cal D}^{(mn)ab}_5{\cal F}^{(n)b\nu 5}\, ,
\end{equation}
\begin{equation}
{\cal D}^{(0)ab}_\mu{\cal F}^{(m)b\mu5}=gf^{abc}\Delta'^{nrm}{\cal F}^{(n)b\mu5}A^{(r)c}_\mu \, ,
\end{equation}
where the object ${\cal D}^{(mn)ab}_{\mu}$ is a sort of covariant derivative, which will be defined in the next subsection. On the other hand, in five dimensions the equations of motion are given by
\begin{equation}
{\cal D}^{ab}_M{\cal F}^{bMN}=0\, ,
\end{equation}
which are equivalent to
\begin{eqnarray}
{\cal D}^{ab}_\mu {\cal F}^{b\mu \nu}+{\cal D}^{ab}_5{\cal F}^{b5\nu}&=&0\, , \\
{\cal D}^{ab}_\mu {\cal F}^{b\mu 5}&=&0 \, .
\end{eqnarray}
Expanding these equations in Fourier series and integrating the fifth dimension, we arrive to the same set of equations of motion derived directly from ${\cal L}_{4YM}$.

\subsection{Standard and nonstandard gauge transformations}
 As emphasized in the introduction, one of our main objectives in this work is to derive the gauge transformations determined by the $\alpha^{(0)a}$ and $\alpha^{(n)a}$ gauge parameters, as it is crucial to quantize the theory. As already commented, we will derive these gauge transformations following more than one method.

 \subsubsection{Four dimensional transformations from standard five dimensional transformations}
 The precise way through which the fields $A^{(0)a}_\mu$ and $A^{(m)a}_\mu$ transform is encoded in the corresponding five dimensional transformation given by Eq.(\ref{TLA}), which can be written as
 \begin{eqnarray}
 \delta {\cal A}^a_\mu (x,y)&=&{\cal D}^{ab}_\mu \alpha^b(x,y)\, , \\
 \delta {\cal A}^a_5 (x,y)&=&{\cal D}^{ab}_5 \alpha^b(x,y)\, .
 \end{eqnarray}
Following exactly the same procedure used in the derivation of the laws of transformation for the curvatures, we obtain
\begin{eqnarray}
\label{gt1}
\delta A^{(0)a}_\mu&=&{\cal D}^{(0)ab}_\mu \alpha^{(0)b}+gf^{abc}A^{(m)b}_\mu \alpha^{(m)c} \, ,\\
\label{gt2}
\delta A^{(m)a}_\mu&=&gf^{abc}A^{(m)b}_\mu \alpha^{(0)c}+{\cal D}^{(mn)ab}_\mu \alpha^{(n)b} \, , \\
\label{gt3}
\delta A^{(m)a}_5&=&gf^{abc}A^{(m)b}_5\alpha^{(0)c}+{\cal D}^{(mn)ab}_5\alpha^{(n)b}\, ,
\end{eqnarray}
where
\begin{eqnarray}
\label{cd1}
{\cal D}^{(mn)ab}_\mu&=&\delta^{mn}{\cal D}^{(0)ab}_\mu-gf^{abc}\Delta^{mrn}A^{(r)c}_\mu \, , \\
\label{cd2}
{\cal D}^{(mn)ab}_5&=&-\delta^{mn}\delta^{ab}\frac{m}{R}-gf^{abc}\Delta'^{mrn}A^{(r)c}_5\, .
\end{eqnarray}
The laws of transformation for the curvatures given by Eqs.(\ref{lt1},\ref{lt2},\ref{lt3}) can be reproduced using these variations of the fields. From these expressions, two types of gauge transformations can be distinguished. One of them is obtained when all the excited KK modes of the gauge parameters are put equal to zero, \textit{i.e.} $\alpha^{(m)a}=0$ for all $m=1,2,\cdots $. In such case,
 \begin{eqnarray}
 \label{sgt1}
\delta A^{(0)a}_\mu&=&{\cal D}^{(0)ab}_\mu \alpha^{(0)b}\, ,\\
\label{sgt2}
\delta A^{(m)a}_\mu&=&gf^{abc}A^{(m)b}_\mu \alpha^{(0)c} \, , \\
\label{sgt3}
\delta A^{(m)a}_5&=&gf^{abc}A^{(m)b}_5\alpha^{(0)c}\, ,
\end{eqnarray}
which show that the zero mode $A^{(0)a}_\mu$ transforms in the standard way of a Yang--Mills theory, whereas the excited KK modes $A^{(n)a}_\mu$ and $A^{(n)a}_5$ transform as matter fields in the adjoint representation of $SU_4(N)$. These are the well known SGT. The consideration of the above scenario is strictly needed in order to recover the standard four dimensional Yang--Mills theory. This suggests to investigate in the same way the role played by the parameters $\alpha^{(m)a}$. As in the previous case, we now put $\alpha^{(0)a}=0$ in the Eqs.(\ref{gt1},\ref{gt2},\ref{gt3}), which leads to
 \begin{eqnarray}
 \label{nsgt1}
\delta A^{(0)a}_\mu&=&gf^{abc}A^{(n)b}_\mu \alpha^{(n)c} \, ,\\
\label{nsgt2}
\delta A^{(m)a}_\mu&=&{\cal D}^{(mn)ab}_\mu \alpha^{(n)b} \, , \\
\label{nsgt3}
\delta A^{(m)a}_5&=&{\cal D}^{(mn)ab}_5\alpha^{(n)b}\, .
\end{eqnarray}
Several comments are in order here. The first point to be noted is that these transformations are much more complicated than the standard ones, as they mix the infinite number of excited modes. We can see that the zero modes $A^{(0)a}_\mu$ do not transform trivially under these NSGT, as suggested in Ref.~\cite{W5}, but they are mapped into excited gauge KK modes, in a way that resembles the standard adjoint representation. Also, we can see that the variations of the KK modes $A^{(n)a}_\mu$ depend on the zero modes $A^{(0)a}_\mu$ through the standard covariant derivative contained in ${\cal D}^{(nm)ab}_\mu$. By contrast, the variation of the scalar KK modes $A^{(n)a}_5$ does not depend on the zero or the excited gauge modes. The mathematical structure of ${\cal D}^{(nm)ab}_\mu$ suggests that the KK modes $A^{(n)a}_\mu$ are gauge fields under this second sort of gauge transformations. As we will see below, the ${\cal D}^{(nm)ab}_\mu$  tensorial structure play a central role in deriving the ghost sector associated with the excited KK modes.

Let us to conclude this part by showing that the scalar fields $A^{(m)a}_5$ can be eliminated altogether via a particular NSGT. Consider a NSGT with infinitesimal gauge parameters given by $\alpha^{(m)a}(x)=(R/m)A^{(m)a}_5$. Then, from Eq.(\ref{nsgt3}), we can see that $A^{(m)a}_5 \to A'^{(m)a}_5=0$. On the other hand, the sole term that contains the $A^{(m)a}_5$ scalar fields is the NSGT invariant object $\frac{1}{2}{\cal F}^{(m)a}_{\mu 5} {\cal F}^{(m)a\mu}_5$, which, in this particular gauge, takes the way:
\begin{equation}
\frac{1}{2}{\cal F}^{(m)a}_{\mu 5} {\cal F}^{(m)a\mu}_5 \to \frac{1}{2}{\cal F}'^{(m)a}_{\mu 5} {\cal F}'^{(m)a\mu}_5=\frac{1}{2}\left(\frac{m}{R}\right)^2A^{(m)a}_\mu
A^{(m)a\mu} \,.
\end{equation}
This result shows that the $A^{(m)a}_5(x)$ scalar fields are in fact pseudo Goldstone bosons.

\subsubsection{Dirac's method}

The SGT and the NSGT can also be derived by using the Dirac's method~\cite{DM}, which allows us to study the phase space constraints to which the gauge system is subject. Once the constraints of the system are known, one constructs the Castellani's gauge generator through which the variations of the fields can be determined. In the Dirac's formalism, one needs to determine the generalized momenta, which are given by
\begin{eqnarray}
\pi^{(0)a}_\alpha&=&\frac{\partial {\cal L}_{4YM}}{\partial \dot{A}^{(0)a}_\alpha}={\cal F}^{(0)a}_{\alpha 0} \, , \\
\pi^{(m)a}_\alpha&=&\frac{\partial {\cal L}_{4YM}}{\partial \dot{A}^{(m)a}_\alpha}={\cal F}^{(m)a}_{\alpha 0} \, , \\
\pi^{(m)a}_5&=&\frac{\partial {\cal L}_{4YM}}{\partial \dot{A}^{(m)a}_5}={\cal F}^{(m)a}_{05} \, ,
\end{eqnarray}
where the dot over the fields denotes velocities. For $\alpha=i$, the above expressions can be solved for the following velocities
\begin{eqnarray}
\dot{A}^{(0)ia}&=&\pi^{(0)a}_i-{\cal D}^{(0)ab}_iA^{(0)b}_0-gf^{abc}A^{(m)b}_iA^{(m)c}_0\, ,\\
\dot{A}^{(m)ia}&=&\pi^{(m)a}_i-{\cal D}^{(mn)ab}_iA^{(n)b}_0-gf^{abc}A^{(m)b}_iA^{(0)c}_0\, ,\\
\dot{A}^{(m)a}_5&=&\pi^{(m)a}_5-{\cal D}^{ (mn)ab}_5A^{(n)b}_0+gf^{abc}A^{(m)b}_5A^{(0)c}_0\, ,
\end{eqnarray}
whereas for $\alpha=0$ one has the primary constraints
\begin{eqnarray}
\label{p1}
\phi^{(1) (0)}_a&\equiv \pi^{(0)a}_0 \approx 0\, , \\
\label{p2}
\phi^{(1)(m)}_a &\equiv \pi^{(m)a}_0 \approx 0\, ,
\end{eqnarray}
where the label $(1)$ stands for primary. On the other hand, the primary Hamiltonian is given by
\begin{equation}
H^{(1)}=\int d^3x {\cal H}^{(1)}\, ,
\end{equation}
with
\begin{equation}
{\cal H}^{(1)}={\cal H}+\lambda^{(0)a}\phi^{(1)(0)}_a+\lambda^{(m)a}\phi^{(1)(m)}_a\, ,
\end{equation}
where $\lambda^{(0)a}$ and $\lambda^{(m)a}$ are Lagrange multipliers. In addition,
\begin{eqnarray}
{\cal H}&=&+\frac{1}{2}\pi^{(0)a}_i\pi^{(0)a}_i+\frac{1}{2}\pi^{(m)a}_i\pi^{(m)a}_i+\frac{1}{2}\pi^{(m)a}_5\pi^{(m)a}_5 \nonumber \\
&&+A^{(0)a}_0{\cal D}^{(0)ab}_i\pi^{(0)b}_i+A^{(m)a}_0{\cal D}^{(mn)ab}_i\pi^{(n)b}_i-A^{(n)b}_0{\cal D}^{(mn)ab}_5\pi^{(m)a}_5 \nonumber \\
&&-gf^{abc}\left(\pi^{(0)a}_iA^{(m)b}_iA^{(m)c}_0 +\pi^{(m)a}_iA^{(m)b}_iA^{(0)c}_0-\pi^{(m)a}_5A^{(m)b}_5A^{(0)c}_0\right)\nonumber \\
&&+\frac{1}{4}\left({\cal F}^{(0)a}_{ij}{\cal F}^{(0)a}_{ij} +{\cal F}^{(m)a}_{ij}{\cal F}^{(m)a}_{ij}+2{\cal F}^{(m)a}_{i5}{\cal F}^{(m)a}_{i5}\right)\, .
\end{eqnarray}
We now demand that the primary constraints satisfy the consistency conditions:
\begin{eqnarray}
\dot{\phi}^{(1)(0)}_a=\{\phi^{(1)(0)}_a(x),H^{(1)}\}\approx 0\, ,\\
\dot{\phi}^{(1)(m)}_a=\{\phi^{(1)(m)}_a(x),H^{(1)}\}\approx 0\, ,
\end{eqnarray}
which leads to secondary constraints, given by
\begin{eqnarray}
\label{s1}
\phi^{(2)(0)}_a&=&{\cal D}^{(0)ab}_i\pi^{(0)b}_i-gf^{abc}\left(\pi^{(m)b}_iA^{(m)c}_i+\pi^{(m)b}_5A^{(m)c}_5 \right)\approx0 \, ,\\
\label{s2}
\phi^{(2)(m)}_a&=&{\cal D}^{(mn)ab}_i\pi^{(n)b}_i-{\cal D}^{(mn)ab}_5\pi^{(n)b}_5-gf^{abc}\pi^{(0)b}_iA^{(m)c}_i \approx 0\, .
\end{eqnarray}
After a tedious algebra, one finds the following relations
\begin{equation}
\label{r1}
\{\phi^{(2)(0)}_a(x),\phi^{(2)(0)}_b(x')\}=gf_{abc}\phi^{(2)(0)}_c(x)\delta(\vec{x}-\vec{x}')\, ,
\end{equation}
\begin{equation}
\label{r2}
\{\phi^{(2)(0)}_a(x),\phi^{(2)(m)}_b(x')\}=gf_{abc}\phi^{(2)(m)}_c(x)\delta(\vec{x}-\vec{x}')\, ,
\end{equation}
\begin{equation}
\label{r3}
\{\phi^{(2)(m)}_a(x),\phi^{(2)(n)}_b(x')\}=gf_{abc}\left(\delta^{mn}\phi^{(2)(0)}_c(x)+\Delta^{mrn}\phi^{(2)(r)}_c \right)\delta(\vec{x}-\vec{x}')\, .
\end{equation}
Since the Poisson's brackets of the primary constraints with the secondary ones also vanish, the constraints of the four dimensional theory are first class constraints, showing thus that excited KK modes $A^{(m)a}_\mu$ are gauge fields.

It is worth noticing that the above set of primary and secondary constraints can be derived directly from the five dimensional theory. At this level, the momenta definition leads to
\begin{equation}
\pi^a_A (x,y)=\frac{\partial {\cal L}_{5YM}}{\partial \dot{A}^{(0)a}_A}={\cal F}^{a}_{A 0}(x,y)\, ,
\end{equation}
so, for $A=0$ the primary constraints arise:
\begin{equation}
\phi^{(1)}_a(x,y)\equiv \pi^{a}_0(x,y) \approx 0\, .
\end{equation}
Notice that, from its definition, the momentum $\pi^a_i(x,y)$ has even parity, whereas $\pi^a_5$ is odd. We will assume that $\pi^a_0$ or, equivalently $\phi^{(1)}_a$, has even parity. The secondary constraints are given by
\begin{equation}
\begin{array}{rclr}
\phi^{(2)}_a&\equiv& {\cal D}^{ab}_I\pi^b_I& \,  \, \, I=i,5 \, \nonumber \\
&=&{\cal D}^{ab}_i\pi^b_i+{\cal D}^{ab}_5\pi^b_5 \approx 0 . \, &
\end{array}
\end{equation}
As before, we expand in Fourier series both the left--hand and right--hand sides of the above expressions and next we use the orthogonality of the trigonometric functions to link the corresponding Fourier modes. For instance, the primary constraints are expanded as follows
\begin{equation}
\frac{1}{\sqrt{2\pi R}}\phi^{(1)(0)}_a(x)+\sum_{m=1}^\infty \frac{1}{\sqrt{\pi R}}\phi^{(1)(m)}_a(x)\cos\left(\frac{my}{R}\right)=\frac{1}{\sqrt{2\pi R}}\pi^{(0)a}_0(x)+\sum_{m=1}^\infty \frac{1}{\sqrt{\pi R}}\pi^{(m)a}_0(x)\cos\left(\frac{my}{R}\right) \, .
\end{equation}
Next, multiply by $1/\sqrt{2\pi R}$, integrate over $y$, and use the orthogonality of the trigonometric functions. The result is the primary constraint given by Eq.(\ref{p1}). Then multiply by $(1/\sqrt{\pi R}) \cos\left(ny/R\right)$, integrate over $y$ and use again the orthogonality of the trigonometric functions. The result is the primary constraint given by Eq.(\ref{p2}). Applying the same procedure to the five dimensional secondary constraints $\phi^{(2)}_a(x,y)$, one reproduces the four dimensional secondary constraints given by Eqs.(\ref{s1},\ref{s2}). Finally, the gauge algebra given by Eqs.(\ref{r1},\ref{r2},\ref{r3}) can be derived directly from the five dimensional counterpart:
\begin{equation}
\{\phi^{(2)}_a(x,y),\phi^{(2)}_b(x',y')\}=g_5f_{abc}\phi^{(2)}_c(x,y)\delta (\vec{x}-\vec{x}')\delta(y-y')\, .
\end{equation}
In this case, once the constraints inside of the Poisson's brackets are expanded in Fourier series, a double integration over the variables $y$ and $y'$ is needed in order to reproduce the relations given by Eqs.(\ref{r1},\ref{r2},\ref{r3}).

The knowing of the constraints is essential to construct the Castellani's gauge generator~\cite{Castellani}, which allows us to determine the gauge transformations of the gauge fields. Now we show that this formalism reproduces both the SGT and the NSGT given by Eqs.(\ref{sgt1},\ref{sgt2},\ref{sgt3}) and Eqs.(\ref{nsgt1},\ref{nsgt2},\ref{nsgt3}), respectively. In our case, such a generator can be written as follows
\begin{eqnarray}
{\cal G}&=&\int d^3z\Big[\left({\cal D}^{(0)ab}_0\alpha^{(0)b}+gf^{abc}A^{(m)b}_0\alpha^{(m)c} \right)\phi^{(1)(0)}_a
-\alpha^{(0)a}\phi^{(2)(0)}_a
\nonumber \\
&&
+\left(
gf^{abc} A^{(m)b}_{0} \alpha^{(0)c}
+ {\cal D}^{(mn)ab}_{0} \alpha^{(n)b}
\right) \phi^{(1)(m)}_{a}
-\alpha^{(m)a}\phi^{(2)(m)}_a\Big]\, ,
\end{eqnarray}
where the functions $\alpha^{(0)a}(x)$ and $\alpha^{(n)a}(x)$ are the gauge parameters, only restricted to be soft. The gauge transformations are obtained by calculating the variations
\begin{eqnarray}
\delta A^{(0)a}_\mu&=&\{A^{(0)a}_\mu,{\cal G}\}\, , \\
\delta A^{(m)a}_\mu&=&\{A^{(m)a}_\mu,{\cal G}\}\, , \\
\delta A^{(m)a}_5&=&\{A^{(m)a}_5, {\cal G}\}\, .
\end{eqnarray}
The calculation of these Poisson's brackets just leads to Eqs.(\ref{gt1},\ref{gt2},\ref{gt3}), which in turn implies the SGT and NSGT given by Eqs.(\ref{sgt1},\ref{sgt2},\ref{sgt3}) and Eqs.(\ref{nsgt1},\ref{nsgt2},\ref{nsgt3}), respectively.

\subsection{Scenario with gauge parameters confined to the brane}
As it has been emphasized through the paper, the BRST formalism introduces the gauge parameters from the outset, \textit{i.e.} at the classical level, as true degrees of freedom, as these ghosts fields have been useful throughout the development of covariant gauge systems quantization. Due to this, it is important to study the physical implications of considering the scenarios that can arise when the gauge parameters propagate in the fifth dimension or are confined to the brane. The former possibility was studied previously. Here, we will study the consequences of assuming that the gauge parameters $\alpha^a$ do not depend on the fifth dimension. Previously, we showed that if the gauge parameters propagate in the fifth dimension, in order to preserve gauge--invariance, one must compactify the theory by considering the curvature as the fundamental objects rather than the fields. In the scenario with the gauge parameters confined in the brane, there are no excited modes for such parameters and no gauge symmetry can arise other than the standard one, \textit{i.e.} that characterized by the SGT studied above. As it is evident from Eqs. (\ref{sgt1},\ref{sgt2},\ref{sgt3}), the gauge fields correspond to the zero modes of the five dimensional gauge field, whereas the excited ones transform as matter fields in the adjoint representation of the group. Indeed, the scenario that will be studied here is entirely determined by the SGT and the four dimensional theory differs from the one derived above, as the corresponding Lagrangian, which we will denote by $\hat{{\cal L}}_{4YM}$, does not coincide with ${\cal L}_{4YM}$.

If the gauge parameters do not propagate in the fifth dimension, the variation of the five dimensional gauge fields given by Eq.(\ref{TLA}) takes the way
\begin{eqnarray}
\delta {\cal A}^a_\mu(x,y)&=&{\cal D}^{ab}_\mu \alpha^b(x)\, , \\
\delta {\cal A}^a_5(x,y)&=&g_5f^{abc}{\cal A}^b_5(x,y)\alpha^c(x)\, .
\end{eqnarray}
The last equation in the above expressions clearly shows that the fields ${\cal A}^a_5$ transform in the adjoint representation of the group even before the compactification. Since in this scenario $SU_5(N)=SU_4(N)\equiv SU(N)$, both the five dimensional theory and the four dimensional one must be governed by the well known SGT. As before, we start from the Lagrangian given by Eq.(\ref{L4}). However, instead of using the Fourier series for the curvatures ${\cal F}^a_{\mu \nu}$ and ${\cal F}^a_{\mu 5}$, given by Eqs.(\ref{ec1},\ref{ec2}), we will use the expressions given in Eqs.(\ref{ec3},\ref{ec4}), which are obtained after expanding in Fourier series the gauge fields ${\cal A}^a_\mu$ and ${\cal A}^a_5$. To clarify why each procedure leads to different four dimensional theories, let us to comment the reason behind this. If the integrand in Eq.(\ref{L4}) is defined by expanding directly in Fourier series the curavatures, one must solve integrals of the way
\begin{equation}
\int^{2\pi R}_0 dy\left(F_0+F_m\cos\left(\frac{my}{R}\right)\right)^2=\pi R(2 F^2_0+F^2_m)\, .
\end{equation}
By contrast, if the integrand is defined by expanding in Fourier series the gauge fields instead of the curvatures, one must solve integrals of the way
\begin{eqnarray}
\int^{2\pi R}_0dy\left(F_0+F_m\cos\left(\frac{my}{R}\right)+F_{mn}\cos\left(\frac{my}{R}\right)\cos\left(\frac{ny}{R}\right)\right)^2&=&\pi R\Big(2F^2_0+F^2_m+2F_0F_{mm}\nonumber \\
&&+\Delta^{mrn}F_mF_{rn}+\Delta^{mnrs}F_{mn}F_{rs}\Big)\, ,
\end{eqnarray}
where the factor with two indices, $F_{mn}$, represents the Fourier coefficients of the non--Abelian part of the curvature ${\cal F}^a_{MN}$. This is why the NSGT violate gauge invariance. Also, this is the reason why gauge invariance is manifest in Abelian theories when one follows this path~\cite{W2}, as in this case there is no mathematical difference between the Fourier series of the curvature and that of the gauge fields. After these considerations, we turn to evaluate the integral (\ref{L4}) by expanding in Fourier series the gauge fields ${\cal A}^a_\mu$ and ${\cal A}^a_5$. After carrying out the integrations and performing some algebraic manipulations, one obtains
\begin{equation}
\hat{{\cal L}}_{4YM}={\cal L}_{4YM}+\Delta {\cal L}\, ,
\end{equation}
where
\begin{equation}
\Delta {\cal L}=\frac{1}{4}g^2 f^{abc}f^{ade}\left(\delta^{rnpq} A^{(n)c}_\nu A^{(q)e \nu}-\delta'^{rnpq} A^{(n)c}_5A^{(q)e}_5 \right)A^{(r)b}_\mu A^{(p)d\mu}\, .
\end{equation}
In the above expression,
\begin{eqnarray}
\delta^{rnpq}&=&\delta^{rn}\delta^{pq}+\Delta^{mrn}\Delta^{mpq}-\Delta^{rnpq}\, ,\\
\delta'^{rnpq}&=&\Delta'^{mrn}\Delta'^{mpq}-\Delta'^{rnpq}\, ,
\end{eqnarray}
with
\begin{eqnarray}
\Delta^{rnpq}&=&\frac{1}{2}\left(\delta^{r,n+p+q}+\delta^{n,r+p+q}+\delta^{p,r+n+q}+\delta^{q,r+n+p}+\delta^{r+n,p+q}+\delta^{r+p,n+q}+\delta^{r+q,n+p}\right)\, ,\\
\Delta'^{rnpq}&=&\frac{1}{2}\left(-\delta^{r,n+p+q}+\delta^{n,r+p+q}-\delta^{p,r+n+q}+\delta^{q,r+n+p}+\delta^{r+n,p+q}-\delta^{r+p,n+q}+\delta^{r+q,n+p}\right)\, .
\end{eqnarray}
Notice that the new term, $\Delta {\cal L}$, is invariant under the SGT but not under the NSGT. It is important to notice that it is not possible to eliminate from the theory the scalar fields $A^{(m)a}_5$, as in this case, where gauge parameters are confined to the brane, the NSGT are absent.

We have shown before that the assumption that the gauge parameters are allowed to propagate in the fifth dimension leads to first class constraints, but not to second class ones. Second class constraints can arise due to the presence of massive vector fields (matter fields or Proca Fields). We now turn to show that this system is subject to both first class and second class constraints. To see this, we need to study the constraints to which is subject the system characterized by the  $\hat{{\cal L}}_{4YM}$ Lagrangian, which is simple to do, indeed, as this Lagrangian differs from ${\cal L}_{4YM}$ only by the  term $\Delta{\cal L}$, which does not contain terms with derivatives. This means that the momenta generated by the $\hat{{\cal L}}_{4YM}$ theory, and therefore the primary constraints, are the same than those induced by ${\cal L}_{4YM}$. The consitency condition on the primary constraint $\phi^{(1)(0)}_a=\pi^{(0)a}_0$, which leads to the secondary constraint $\phi^{(2)(0)}_a$, nothing new introduce because $\Delta {\cal L}$ does not depend on $A^{(0)a}_0$, the canonical conjugate of $\pi^{(0)a}_0$, and therefore, the secondary constraint $\phi^{(2)(0)}_a$ remains unchanged. On the other hand, the consistency condition on the primary constraint $\phi^{(1)(m)}_a=\pi^{(m)a}_0$ produces some changes, as the new term $\Delta {\cal L}$ does depend on $A^{(m)a}_0$, the canonical conjugate of $\pi^{(m)a}_0$. Taking into account that the primary Hamiltonians of the two theories are related in a simple way,
\begin{equation}
\hat{H}^{(1)}=H^{(1)}-\int d^3x \Delta {\cal L}\, ,
\end{equation}
it is easy to determine the way of the new secondary constraint. In fact, the consistency condition for $\phi^{(1)(m)}_a$
\begin{equation}
\dot{\phi}^{(1)(m)}_a=\{\phi^{(1)(m)}_a(x),\hat{H}^{(1)}\}\approx 0\, ,
\end{equation}
leads to
\begin{equation}
\hat{\phi}^{(2)(m)}_a=\phi^{(2)(m)}_a+g^2f^{abc}f^{bde}\left(\delta^{mnpq}A^{(n)c}_\mu A^{(q)e\mu}-\delta^{mnpq}A^{(n)c}_5A^{(q)e}_5 \right) A^{(p)d}_0 \, .
\end{equation}
It is easy to see that the consistency condition on $\hat{\phi}^{(2)(m)}_a$ determine the Lagrange multipliers $\lambda^{(m)a}$ appearing in ${\cal H}^{(1)}$ because of
\begin{equation}
\{\hat{\phi}^{(2)(m)}_a(x),\phi^{(1)(n)}_b(x')\}\neq 0\, .
\end{equation}
This in turn implies that the constraints $\phi^{(1)(m)}_a$ and $\hat{\phi}^{(2)(m)}_a$ are second class constraints. Since the relations given by Eqs.(\ref{r1},\ref{r2}) remain unchanged, the constraints $\phi^{(1)(0)}_a$ and $\phi^{(2)(0)}_a$ are first class constraints. This is just the gauge structure of a theory that contains both gauge fields (the zero modes $A^{(0)a}_\mu$) and Proca fields (the excited modes $A^{(m)a}_\mu$).

As already commented, the approach followed in the literature when deriving the four dimensional theory is the one presented here, although it should be stressed that in previous works the gauge parameters are taken as living in the bulk, whereas we demand that they are confined to the brane, which is necessary in order to preserve gauge invariance. However, the corresponding Lagrangian must coincide in both schemes because the only difference is the absence in a case of excited KK modes of the gauge parameters and the presence of them in the other scenario. The way in which the four dimensional Lagrangian is derived is the same in both schemes. Then, it is worth comparing our Lagrangian $\hat{{\cal L}}_{4YM}$ with those obtained in the literature.

The nonabelian model analyzed in this work is interesting and important as it can be part of extra dimensional Standard Model extensions and employed to perform phenomenological calculations \cite{FMNRT1,FMNRT2}. This five dimensional Lagrangian and its transition to a four dimensional effective version, reached through the compactification process, have been studied and some results have been reported in the literature. In reference \cite{Dienesetal} the authors analyzed some aspects about extra dimensional models. In connection with our work, in that time they developed an abelian gauge model with one extra dimension, which they compactified on the orbifold $S^{1}/Z_{2}$. They obtained a set of gauge transformations for such a model and argued in favor of the renormailzability of it when truncated to a certain order of the Fourier series. Later, they commented that the same procedure carried out by them could be repeated for the nonabelian case, and that the results concerning the renormalizability of this more elaborated model followed straightforwardly. They did not present any expression for the effective four dimensional Lagrangian nor a precise explanation of how they performed the integration of the extra dimension. Nonetheless, they offered a set of gauge transformations, at the four dimensional level, which do not coincide with the ones that we obtained. Other related works are \cite{W1} and \cite{Chivukulaetal}, where the five dimensional Yang-Mills theory was compactified and the Lagrangian Fourier--expanded by taking the gauge fields as the fundamental objects instead of the curvatures. The expressions exhibited in both papers, concerning the structure of the four dimensional effective Lagrangian, match our results for the case in which the gauge parameters are constrained to the brane. In both papers the authors did not study the gauge transformations that rule the four dimensional effective Lagrangian. The authors in \cite{W2} studied some extra dimensional extensions of the Standard Model, and also discussed some aspects concerning an extra dimensional Yang-Mills theory, for which they defined a nonabelian Lagrangian as usual, and considered gauge--fixing and Fadeev--Popov terms. They compactified the model on the orbifold $S^{1}/Z_{2}$ and offered Feynman rules for the four dimensional effective theory. In such a work, the expression for the 4D effective Lagrangian was not explicitly shown, but the presence of the factors $\Delta^{nmkl}$ and $\Delta '^{nmkl}$ in their Feynman rules suggest that they expanded directly the gauge fields instead of the curvatures. Another interesting element in this paper is the exhibition of the gauge transformations corresponding to the four dimensional effective Lagrangian. We have compared such transformation laws with those that we presented in this paper and found a perfect agreement in the case of the variations of the fields $A^{(0)a}_{\mu}$ and $A^{(n)a}_{5}$. By contrast, we encountered some differences when we analyzed the remaining transformations: $\delta A^{(n)a}_{\mu} = our \hspace{0.1cm} result - \sqrt{2}gf^{abc} A^{(0)b}_{\mu} \alpha^{(n)c}$. Concerning our work, in \cite{chinos} the authors studied, among other things, a five dimensional electroweak Standard Model Extension. The Yang-Mills sector that they considered for such a model differs from ours only because they included contact terms. They performed an orbifold compactification on $S^1/Z_2$ and then expanded the fields in sine and cosine series. It is remarkable that in this paper the authors expanded the curvatures and obtained the analogous four dimensional objects. They did not emphasize their procedure nor said any word about integrating curvatures instead of fields in order to preserve gauge invariance, but they showed the four dimensional expressions for the curvatures after introducing a classical background field, this in the context of the background field method. We have compared our expressions with those exhibited by these authors and have found agreement. Finally, they did not derive the gauge transformations governing the four dimensional effective Lagrangian. Another interesting reference is \cite{W5}, where some aspects related to the gauge structure of the usual five dimensional SU(N) model were analyzed. The author incorporated two brane kinetic terms to the nonabelian Lagrangian and compactified the model on the orbifold $S^{1}/Z_{2}$, so, except for the brane kinetic terms, the initial considerations taken by him are the same than in our case. He showed an explicit expression for the four dimensional effective Lagrangian, but he did not talk about the precise procedure followed by him in order to obtain such an expression. The structure of his result suggests that he employed the same method than us, that is, he integrated the fifth dimension over the curvatures rather than over the gauge fields. However, our result do not exactly coincide with that presented by him. In fact, when one neglects the brane kinetic terms, his Lagrangian can be written as ${\cal L}_{eff} = our \hspace{0.3cm} result - g^{2} f^{abc} f^{ade} A^{(m)b}_{\mu}A^{(m)c}_{\nu}A^{(k)d\mu}A^{(k)e\nu}$. This author also presented in his paper a set of gauge transformations for the gauge fields, which he divided into two types: the SGT, defined by the zero order parameters, and those determined by all the other parameters, which are the analogous of our NSGT. The SGT found by this author are in perfect agreement with our results. Nonetheless, there is a remarkable difference in the case of the remaining variations, for the expressions found in this reference indicate that the transformations for the zero-mode fields vanish, that is, there is no variation of the zero-mode fields under such transformations. Contrastingly, in the NSGT that we report the transformations corresponding to the zero modes are not zero, but something similar to adjoint transformations that mix different KK modes.

\section{Quantization}
\label{Q}The purpose of this section is to carry out the quantization of the four dimensional KK gauge theory discussed in the previous section. We will focus on the scenario in which the gauge parameters propagate in the fifth dimension. The classical theory to be quantized is characterized by the ${\cal L}_{4YM}$ Lagrangian, which is invariant under both the SGT and the NSGT of $SU_4(N)$. In particular, we are interested in quantizing the excited KK gauge modes $A^{(m)a}_\mu$, as our main objective is to investigate the quantum loop fluctuations of these gauge fields on light Green functions, $\langle0|T(A^{(0)a}_\mu(x_1)A^{(0)b}_\nu(x_2)\cdots |0\rangle$. This means that it is only necessary a fixation of the gauge with respect to the NSGT. We will do this by implementing a gauge--fixing procedure that is covariant under the SGT. Our discussion will be based on the BRST symmetry.

\subsection{The proper solution of the master equation}
Classically, the BRST symmetry arises naturally within the context of the Batalin--Vilkovisky formalism~\cite{BV}, also known as field--antifield formalism~\cite{PR}. To clarify our presentation, let us to present a brief discussion on this formalism. Although the following discussion is rather general, we will focus on the properties of Yang--Mills systems. The starting point is the introduction of an antifield for each field present in the theory. It is assumed that the dynamical degrees of freedom of the gauge system comprise, besides the gauge fields, the ghost ($C^a$), antighost ($\bar{C}^a$), and auxiliary ($B^a$) fields. The original action, which will be denoted by $S_0$, is a functional of the gauge fields only, but this configuration is extended to include the ghost fields because they are necessary to quantize the theory. A ghost field for each gauge parameter is introduced. The ghost fields have opposite statistics to that of the gauge parameters. To gauge-fix and quantize the theory, it is necessary to introduce the so-called trivial pairs, namely the antighost and auxiliary fields. We let $\Phi^A$ run over all the fields. For each $\Phi^A$, an antifield $\Phi^*_A$ is introduced, with opposite statistics to $\Phi^A$ and a ghost
number equal to $-gh(\Phi^A)-1$, where $gh(\Phi^A)$ is the ghost number of $\Phi^A$. It is $0$  for matter, gauge and auxiliary fields, $+1$ for ghosts and $-1$ for antighosts. In this extended configuration space, a symplectic structure, called antibracket, is introduced through left and right differentiation, defined for the two functionals $F$ and $G$ as
\begin{equation}
\left(F,G\right)=\frac{\partial_R F}{\partial \Phi^A}\frac{\partial_L G}{\partial \Phi^*_A}-\frac{\partial_R F}{\partial \Phi^*_A}\frac{\partial_L G}{\partial \Phi^A}\, .
\end{equation}
In particular, the fundamental atibrackets are given by
\begin{eqnarray}
\left(\Phi^A, \Phi^*_B\right)&=&\delta^A_B \, \\
\left(\Phi^A, \Phi^B\right)=&0&=\left(\Phi^*_A, \Phi^*_B\right)\, ,
\end{eqnarray}
so the antifield $\Phi^*_A$ is canonically conjugate to the field $\Phi^A$ in this sense. The extended action is a bosonic functional of the fields and antifields, $S[\Phi,\Phi^*]$, with zero ghost number, which satisfies the master equation defined by
\begin{equation}
\left(S\, ,S\right)=2\frac{\partial_RS}{\partial \Phi^A}\frac{\partial_L S}{\partial \Phi^*_A}=0\, .
\end{equation}
The extended action is the generator of the BRST transformations:
\begin{eqnarray}
\label{BRSTT}
\delta_B\Phi^A&=&\left(\Phi^A,S\right)\, ,\\
\delta_B\Phi^*_A&=&\left(\Phi^*_A,S\right)\, .
\end{eqnarray}
Notice that $S$ is BRST--invariant due to the master equation. The solutions of the master equation which are of physical interest are those called proper solutions~\cite{PR}. A proper solution must make contact with the initial theory, which means to impose the following boundary condition on $S$:
\begin{equation}
S[\Phi,\Phi^*]|_{\Phi^*=0}=S_0[\phi]\, ,
\end{equation}
where $\phi$ runs only over the original fields. The proper solution can be expanded in power series in antifields,
\begin{equation}
S[\Phi,\Phi^*]=S_0[\phi]+(\delta_B \Phi^A)\Phi^*_A+\cdots,
\end{equation}
in which all the gauge--structure tensors characterizing the gauge system appear. In this sense, the proper solution $S$ is the generating functional of the gauge-structure tensors. $S$ also generates the gauge algebra through the master equation. So, classically, a gauge system is completely determined when the proper solution $S$ is established and the master equation is calculated, which yields the relations that must be satisfied by the gauge-structure tensors. In general, the variations of the fields $(\delta_B \Phi^A)$ are not known from outset, so the most general solution with gauge--structure tensors instead of these explicit variations must be proposed and use the master equation to determine them. This approach could be followed in determining the tensorial structure of the NSGT. Although this is feasible, we will follow an less bothersome alternative, which consists in using the well known proper solution for Yang--Mills theories formulated in a four dimensional spacetime, which can automatically be translated to five dimensions. The compactification of the fifth dimension and its integration must lead to a four dimensional proper solution of the master equation. We will see below that this is indeed the case. In particular, there arise just the same  gauge--structure tensors of the SGT and NSGT derived before by following other methods.

The proper solution of the master equation for Yang--Mills theories in five dimensions can be written as follows
\begin{equation}
S=\int d^4x \int dy\left(-\frac{1}{4}{\cal F}^a_{MN}{\cal F}^{MN}_a+{\cal A}^*_{Ma}{\cal D}^{abM}{\cal C}^b+\frac{1}{2}g_5f^{abc}{\cal C}^*_c{\cal C}^b{\cal C}^a+\bar{{\cal C}}^{*a}B_a  \right)\, ,
\end{equation}
which is a trivial generalization of the corresponding solution in four dimensions. Notice that this proper solution satisfies the boundary condition $S|_{\Phi^*=0}=S_0$, with $S_0$ given by Eq.(\ref{S0}). The term corresponding to the original action, $S_0$, is treated in the same way as Sec. \ref{L4YM}. This leads to an action with four dimensional Lagrangian ${\cal L}_{4YM}$. The remaining terms are also expanded in Fourier series. A parity for the antifield identical to its corresponding field is assumed. The ghost fields have the same parity of the gauge parameters. An even parity for the antifield of the antighost field is assumed, as the presence of the zero mode of the antighost fields is needed in order to recover the four dimensional proper solution of the master equation. By the same token, an even parity for the auxiliary fields $B_a$ is assumed. The variations of the gauge fields in five dimensions are recognized as fundamental objects in the same sense than the curvatures. So, in the second term in the above expression, the Fourier expansion is realized on ${\cal A}^*_{Ma}(\delta {\cal A}^{Ma})$ instead of ${\cal A}^*_{Ma}({\cal D}^{abM}{\cal C}^b)$. After doing this, one obtains
\begin{eqnarray}
S&=&\int d^4x\Big[{\cal L}_{4YM}+A^{(0)*}_{\mu a}{\cal D}^{(0)ab\mu}C^{(0)b}+\bar{C}^{(0)*a}B^{(0)}_a+\frac{1}{2}gf^{abc}C^{(0)*}_cC^{(0)b}C^{(0)a}\nonumber \\
&&+A^{(m)*}_{\mu a}{\cal D}^{(mn)ab\mu}C^{(n)b}-A^{(m)*}_{5a}{\cal D}^{(mn)ab}C^{(n)b}+\bar{C}^{(m)*a}B^{(m)}_a+\frac{1}{2}gf^{abc}C^{(0)*}_cC^{(m)b}C^{(m)a}\nonumber \\
&&+\frac{1}{2}f^{abc}C^{(m)*}_c\left(C^{(0)b}C^{(m)a}+C^{(0)a}C^{(m)b}+\Delta^{mrn}C^{(r)b}C^{(n)a} \right)
\Big]\, .
\end{eqnarray}
Notice that if all the excited KK modes are deleted, the well known proper solution for Yang--Mills theories is reproduced~\cite{PR}. $S$ also is a proper solution of the master equation. On the other hand, a straightforward calculation of Eq.(\ref{BRSTT}) allows us to recover the SGT and NSGT given by Eqs. (\ref{sgt1},\ref{sgt2},\ref{sgt3}) and (\ref{nsgt1},\ref{nsgt2},\ref{nsgt3}), respectively.

\subsection{$SU_4(N)$--covariant gauge--fixing procedure}
Having studied the classical structure of the KK theory, we turn now to carry out its quantization, for which one starts by fixing the gauge, since the extended action is degenerate, and hence cannot be quantized directly. Furthermore, the antifields do not represent true degrees of freedom, so they must be removed before quantizing the theory. They cannot be just set to zero, since $S_0$ is degenerate. However, one can remove the antifields instead through a nontrivial procedure and, at the same time, lift the degeneration of the theory. The antifields can be eliminated by introducing a fermionic functional of the fields, $\Psi[\Phi]$, with ghost number $-1$, such that
\begin{equation}
\Phi^*_A=\frac{\partial \Psi}{\partial \Phi^A}\,.
\end{equation}
Note that it is not necessary to distinguish between left- and right-differentiation. In defining a gauge-fixing procedure, the presence of the trivial pairs, $\bar{C}^a$  and $B_a$, is necessary since the only fields with ghost number $-1$ are precisely the antighosts. In the case we are interested with, we only need to remove the degeneration with respect to the NSGT, so we will introduce a Fermionic funtional $\Psi_{NSGT}$ that allows us to remove the excited KK modes of the antifields via the relation
\begin{equation}
\label{EAF}
\Phi^{(m)*}_A=\frac{\partial \Psi_{NSGT}}{\partial \Phi^{(m)A}}\, .
\end{equation}
We introduce the following fermionic funtional
\begin{equation}
\Psi_{NSGT}=\int d^4x\bar{C}^{(m)a}\left(f^{(m)a}+\frac{\xi}{2}B^{(m)a}+gf^{abc}\Delta^{mrn}\bar{C}^{(r)b}C^{(n)c} \right)\, ,
\end{equation}
where $\xi$ is the gauge parameter and $f^{(m)a}$ represents bosonic gauge--fixing functions, which will be conveniently defined below. From this expression and Eq.(\ref{EAF}), one finds
\begin{eqnarray}
A^{(n)*}_{\mu b}&=&\frac{\partial f^{(m)a}}{\partial A^{(n)b}_\mu}\bar{C}^{(m)a}\, ,\\
A^{(n)*}_{5 b}&=&\frac{\partial f^{(m)a}}{\partial A^{(n)b}_5}\bar{C}^{(m)a}\, ,
\end{eqnarray}
\begin{eqnarray}
C^{(m)*}_a&=&gf^{abc}\Delta^{mrn}\bar{C}^{(r)b}\bar{C}^{(n)c} \, , \\
\bar{C}^{(m)*}_a&=&f^{(m)a}+\frac{\xi}{2}B^{(m)a}+2gf^{abc}\Delta^{mrn}\bar{C}^{(r)b}C^{(n)c}\, .
\end{eqnarray}
Once used these relations to eliminate the antifields in $S$, one obtains the gauge--fixed action $S_{\Psi_{NSGT}}$,
\begin{eqnarray}
S_{\Psi_{NSGT}}&=&\int d^4x\Big[{\cal L}_{4YM}+A^{(0)*}_{\mu a}{\cal D}^{(0)ab\mu}C^{(0)b}+\bar{C}^{(0)*a}B^{(0)}_a\nonumber \\
&&+\frac{1}{2}gf^{abc}C^{(0)*}_c\left(C^{(0)b}C^{(0)a}+C^{(m)a}C^{(m)b} \right)\nonumber \\
&&+\bar{C}^{(m)c}\, \frac{\partial f^{(m)c}}{\partial A^{(n)a}_\mu}\, {\cal D}^{(nr)ab\mu}\, C^{(r)b} -\bar{C}^{(m)c}\, \frac{\partial f^{(m)c}}{\partial A^{(n)a}_5}\, {\cal D}^{(nr)ab}_5\, C^{(r)b}\nonumber \\
&&+\frac{\xi}{2}B^{(m)}_aB^{(m)}_a+B^{(m)}_a\left(f^{(m)a}+2gf^{abc} \Delta^{mrn}\bar{C}^{(r)b}C^{(n)c} \right) \nonumber \\
&&+\frac{1}{2}g^2f^{abc}f^{cde}\Delta^{mpq} \bar{C}^{(p)d}\bar{C}^{(q)e}\left(C^{(0)b}C^{(m)a}+C^{(0)a}C^{(m)b}+\Delta^{mrn}C^{(r)b}C^{(n)a}  \right) \Big]\, .
\end{eqnarray}
Notice that this action is still degenerate with respect to the SGT, unless we introduce terms in $f^{(m)a}$ that breaks explicitly this symmetry. We have proceeded in this way because we are interested only in investigating the loop effects of new physics characterized by the excited KK modes on light Green functions, as this class of effects will be of great importance in future experiments. We will preserve the gauge--invariance with respect to the SGT of $S_{\Psi_{NSGT}}$ by introducing gauge--fixing functions $f^{(m)a}$ that transform covariantly under the SGT. On the other hand, since the auxiliary fields $B^{(m)}_a$ do not propagate and appear quadratically in the action, they can be integrated out in the generating functional. Their integration is equivalent to use directly the equations of motion, given by
\begin{equation}
B^{(m)}_a=-\frac{1}{\xi}\left(f^{(m)a}+2gf^{abc}\Delta^{mrn}\bar{C}^{(r)b}C^{(n)c} \right)\, .
\end{equation}
Once eliminated these fields, one can write the following effective Lagrangian
\begin{equation}
{\cal L}_{eff}={\cal L}_{4YM}+{\cal L}_{GF}+{\cal L}^1_{FPG}+{\cal L}^2_{FPG}\, ,
\end{equation}
where ${\cal L}_{GF}$ is the gauge--fixing term, given by
\begin{equation}
{\cal L}_{GF}=-\frac{1}{2\xi}f^{(m)a}f^{(m)a}\, ,
\end{equation}
whereas ${\cal L}^{(1,2)}_{FPG}$ represent the Faddeev--Popov ghost terms, which are given by
\begin{equation}
{\cal L}^1_{FPG}=\bar{C}^{(m)c}\left(\frac{\partial f^{(m)c}}{\partial A^{(n)a}_\mu}{\cal D}^{(nr)ab \mu}- \frac{\partial f^{(m)c}}{\partial A^{(n)a}_5}{\cal D}^{(nr)ab}_5 \right)C^{(r)b}-\frac{1}{\xi}gf^{abc}\Delta^{mrn}f^{(m)a}\bar{C}^{(r)b}C^{(n)c}\, ,
\end{equation}
\begin{equation}
{\cal L}^2_{FPG}=\frac{1}{2}g^2f^{abc}f^{cde}\Delta^{mpq}\bar{C}^{(p)d}\bar{C}^{(q)e}\left(C^{(0)b}C^{(m)a}+C^{(0)a}C^{(m)b}+\Delta^{mrn}C^{(r)b}C^{(n)a}\,  \right) .
\end{equation}

As already commented, it is desirable to preserve the gauge--invariance of ${\cal L}_{eff}$ under the SGT of $SU_4(N)$. This requires the introduction of gauge--fixing functions $f^{(m)a}$ that transform covariantly under this group. In accordance with this, we introduce the following nonlinear gauge--fixing funtions
\begin{equation}
f^{(m)a}={\cal D}^{(0)ab}_\mu A^{(m)b\mu}-\xi \frac{m}{R}A^{(m)a}_5\, ,
\end{equation}
which, as it is evident, transform under the adjoint representation of $SU_4(N)$. When these gauge--fixing functions are introduced in the gauge--fixing and Faddeev--Popov ghost terms, one obtains
\begin{equation}
\label{GF}
{\cal L}_{GF}=-\frac{1}{2\xi}\left({\cal D}^{(0)ab}_\mu A^{(m)b\mu}\right)\left({\cal D}^{(0)ac}_\nu A^{(m)c\nu}\right)+m_mA^{(m)a}_5\left({\cal D}^{(0)ab}_\mu A^{(m)b\mu}\right)-\frac{1}{2}\xi m^2_m A^{(m)a}_5A^{(m)a}_5\, ,
\end{equation}
which clearly is invariant under the SGT of $SU_4(N)$. Note that an unphysical mass, $\sqrt{\xi}m_m=\sqrt{\xi}(m/R)$, for the pseudo Goldstone bosons $A^{(m)a}_5$  has been generated. On the other hand, the ${\cal L}^1_{FPG}$ Lagrangian can be written as follows
\begin{equation}
{\cal L}^1_{FPG}=\bar{C}^{(m)c}\left({\cal D}^{(0)ac}_\mu\, {\cal D}^{(mn)ab\mu}+\xi m_m\, {\cal D}^{(mn)cb}_5  \right)C^{(m)b}-\frac{1}{\xi}gf^{abc}\Delta^{mrn}f^{(m)a}\bar{C}^{(r)b}C^{(n)c}\, .
\end{equation}
We can write this more explicitly by using the definitions of ${\cal D}^{(mn)ab}_\mu$ and ${\cal D}^{(mn)ab}_5$ given in Eqs.(\ref{cd1},\ref{cd2}):
\begin{eqnarray}
\label{FPG}
{\cal L}^1_{FPG}&=&\bar{C}^{(m)b}\left({\cal D}^{(0)ab}_\mu {\cal D}^{(0)ac\mu}\right)C^{(m)c}-\xi m^2_m\bar{C}^{(m)a}C^{(m)a}\nonumber -gf^{abc}\left[\Delta^{mrn}\bar{C}^{(m)d}\left({\cal D}^{(0)ad}_\mu A^{(r)c\mu} \right)C^{(n)b} 
\right.
\\
&&
-\frac{1}{\xi}\Delta^{mrn}\bar{C}^{(r)c}\left( {\cal D}^{(0)ad}_\mu A^{(m)d\mu} \right) C^{(n)b}
\left.
+\xi m_m \Delta'^{mrn}\bar{C}^{(m)a}A^{(r)c}_5C^{(n)b} 
-m_m \Delta^{mrn}\bar{C}^{(r)a}A^{(m)c}_5 C^{(n)b}
\right]\,.
\end{eqnarray}
Notice that this Lagrangian is invariant under the SGT. Also, notice that unphysical masses for the excited KK ghost--antighost fields have been generated.

An important consequence of our gauge fixing procedure is the elimination of unphysical vertices, which greatly simplifies loop calculations. In fact, we can see that bilinear and trilinear couplings of the form $A^{(m)a}_\mu A^{(n)b}_5$ and $A^{(0)a}_\mu A^{(m)b}_\nu A^{(n)c}_5$ disappear when the terms $(1/2){\cal F}^{(m)a}_5{\cal F}^{(m)a}_5$ and ${\cal L}_{GF}$ are summed together:
\begin{eqnarray}
\frac{1}{2}{\cal F}^{(m)a}_{\mu 5}{\cal F}^{(m)a\mu}_5 + {\cal L}_{GF}&=&m_m\left[ A^{(m)a}_5\left({\cal D}^{(0)ab}_\mu A^{(m)b\mu }\right)+A^{(m)a\mu}
\left({\cal D}^{(0)ab}_\mu A^{(m)b}_5 \right)\right]+\cdots \nonumber \\
&=&m_m\partial_\mu(A^{(m)a}_5A^{(m)b\mu})+\cdots\, .
\end{eqnarray}
By contrast with the conventional linear $R_\xi$ gauges~\cite{LG}, which do not modify the vertices of the theory and explicitly break gauge invariance, in the nonconventional quantization schemes~\cite{Fujikawa,HT,BFM,PT,PRPT,CG331}, as the one presented here, a sort of gauge invariance remains at the quantum level. For instance, the nonlinear gauge--fixing procedure introduced by Fujikawa~\cite{Fujikawa} to define the propagator of the $W^\pm$ weak gauge boson is covariant under the electromagnetic group~\cite{HT}. Although conventional quantization schemes~\cite{LG} are appropriate to calculate $S$--matrix elements, they give rise to ill-behaved off shell Green functions that may display inadequate properties such as a nontrivial dependence on the gauge-fixing parameter, an increase larger than the one observed in physical amplitudes at high energies, and the appearance of unphysical thresholds. It would be interesting if one was able to study the sensitivity to radiative corrections of Green functions without invoking some particular $S$-matrix element. Behind this are the concepts of gauge invariance and gauge independence, which are essential ingredients of the gauge systems. Although the former plays a fundamental role to define the classical action, it does not survive to quantization, as one must invariably invoke an appropriate gauge--fixing procedure to define the quantum action. At the quantum level, the theory is governed by a remnant of the original classical~\cite{PR} BRST symmetry, which is the one first discovered by  Becchi--Rouet--Stora--Tyutin~\cite{BRST}. The generating functional constructed with this class of linear gauges, generates Green functions satisfying the Slavnov--Taylor identities instead of the simpler ones that would exist if the quantum action was gauge--invariant. The preservation of some sort of gauge invariance at the level of the quantum action is the main feature of nonconventional quantization schemes. The most popular are the Background Field Method~\cite{BFM} and the Pinch Technique~\cite{PT}. In the former, each gauge field $A^a_\mu$ is decomposed into a quantum, $Q^a_\mu$, and a classical, $\hat{A}^a_\mu$, parts: $A^a_\mu \rightarrow \hat{A}^a_\mu+Q^a_\mu$. While the effective quantum action is defined through the path integral on the $Q^a_\mu$ fields, the classical fields $\hat{A}^a_\mu$ play the role of sources with respect to which are derived the vertex functions. Due to this, it is only necessary to introduce a gauge--fixing procedure for the quantum fields $Q^a_\mu$ and thus the resultant quantum theory is invariant under gauge transformations of the background fields $\hat{A}^a_\mu$. The Green functions derived in this context satisfy simple (QED--like) Ward identities, which are well--behaved because they contain less unphysical information in comparison with those that arise from the conventional quantization methods. However, it is worth stressing that they are still dependent on the gauge parameter $\xi_Q$ that characterizes the gauge-fixing scheme used for the quantum fields, and so there is no gauge independence. Although gauge dependent, one expects that these Green functions provide us information quite near to the physical reality. At the present, there is still no known mechanism that allows to construct a quantum action from which can be derived both gauge--invariant and gauge--independent Green functions, although the Pinch Technique is a diagrammatic method meant for this purpose~\cite{PT}. This method consists in constructing well--behaved Green functions of a given number of points by combining some individual contributions from Green functions of equal and higher number of points, whose Feynman rules are derived from a conventional effective action or even from a nonconventional scheme. Out of the scope of the Background Field Method and the Pinch Technique, it is still possible to introduce gauge invariance with respect to a subgroup of a given theory. This scheme is particularly useful to assess the virtual effects of heavy gauge bosons lying beyond the Fermi scale on the SM Green functions in a $SU_L(2)\times U_Y(1)$--covariant manner, in which case it is only necessary to introduce a quantization scheme for the heavy fields since the SM fields would only appear as external legs. A scheme of this class was proposed by one of us some few years ago~\cite{CG331} to investigate the loop effects of new heavy gauge bosons predicted by the so--called $331$ models~\cite{PF} on the off shell $W^-W^+\gamma$ and $W^-W^+Z$ vertices. The virtual effects of heavy gauge bosons on light Green functions studied in Ref.\cite{CG331} are predicted by a theory based in the $SU_L(3)\times U_X(1)$ gauge group~\cite{PF} and the gauge--fixing functions introduced transform in the fundamental representation of the usual electroweak group $SU_L(2)\times U_Y(1)$. The gauge--fixing procedure introduced in the present work for the excited KK gauge modes has some similitude with both the Background Field Method and the one of Ref.~\cite{CG331}, as in all of them a sort of nonabelian gauge invariance is maintained at the quantum level. In the Background Field Method, the gauge invariance of the original group is maintained, but with respect to the classical background fields. On the other hand, in the scheme introduced in Ref.\cite{CG331}, gauge invariance is maintained but only with respect to a subgroup of the original group. In the case analyzed in the present work, two types of gauge transformations can be identified, one determined by the zero modes of the gauge parameters $\alpha^{(0)a}$, which was called SGT, and other identified with the excited modes of these parameters, called NSGT. Our gauge--fixing procedure for the excited KK modes is covariant under the SGT of the four dimensional gauge group $SU_4(N)$.

So far there exist different proposals related to the issue of the gauge--fixing of the extra dimensional Yang--Mills theory~\cite{W2,Chivukulaetal,chinos,W5}, some of them invoking the fixation at the five dimensional level, which is not the case in reference~\cite{Chivukulaetal}, where the authors propounded a four dimensional gauge-fixing Lagrangian that has a similar structure to ours, but that is essentially different. Another gauge-fixing scheme for the four dimensional effective model was given in~\cite{W5}, where the author worked within the context of the background field method. As this author recognized two types of gauge transformations, he introduced two types of gauge--fixing functions, one concerning the zero--modes and the others directed to the KK excitations. In the case of the KK excitations, the gauge--fixing functions given in this reference have a very similar structure to the ones proposed by us, but they are different because the former include covariant derivatives with respect to the background gauge fields, whereas we have not even done such a division, originated in the background field method. The reference~\cite{W2} propounds a gauge-fixing approach at the five dimensional level that has been widely used in the literature, as it leads to a proper definition of the propagators for the KK-modes through the convenient cancellation of some bilinear couplings. We have obtained, by means of a Fourier analysis, the four dimensional expressions of the gauge fixing functions given by these authors and verified that they do not match our proposal. This dissimilitude appears because the gauge-fixing Lagrangian defined by these authors involves only simple derivatives operating on fields and no more terms that complete covariant derivatives at the four dimensional level. An interesting point in reference~\cite{chinos} is the employing of a gauge-fixing scheme, which the authors introduced at the five dimensional spacetime level. They divided the fields into background fields and quantum fluctuations, within the context of the background field method, before the Fourier expansions and the integration of the extra dimension, and fixed the gauge differently for these two kinds of fields. In the case of the background fields, they used the unitary gauge, while for the quantum fluctuations they employed a sort of $R_{\xi}$ gauge. They expanded their gauge fixing functions in Fourier series and obtained gauge-fixing conditions for both the zero-mode and the excited modes. The structure of the gauge-fixing functions that these authors obtained for the KK modes include our fixation condition, but they have additional terms which mix KK-mode fields. It is worth emphasizing that within our scheme the fixation of the gauge for the zero modes and that for the KK modes can be performed independently of each other, which is a possibility implicitly present in the separation of the gauge transformations parameters into two types. In fact, in this analysis we have gauge-fixed only the excited modes, while leaving the gauge invariance with respect to the SGT. We chose to fix only the KK excitations as such procedure is crucial to properly study the one-loop quantum contributions of this extra dimensional model to light Green functions. On the other hand, the gauge fixation for the zero modes can be accomplished as usually done for the standard four dimensional Yang-Mills theory or by another scheme, such as e.g. the Background Field Method~\cite{BFM}.

We would like to summarize the main result of this section by displaying the Lagrangian that links up heavy physics with light physics, \textit{i.e.} the Lagrangian that describes the couplings among excited KK modes and zero modes. Such a connection is given by the term ${\cal L}^{(0)(n)}\left(A^{(0)a}_\mu,A^{(n)a}_\mu,A^{(n)a}_5\right)$ of the Lagrangian given in Eq.(\ref{eq1}). This Lagrangian, which receives contributions from Eqs.(\ref{lag},\ref{GF},\ref{FPG}), is made of five pieces that are separately invariant under the SGT:
\begin{eqnarray}
{\cal L}^{(0)(n)}\left(A^{(0)a}_\mu,A^{(n)a}_\mu,A^{(n)a}_5\right)&=&-\frac{1}{2}gf^{abc}F^{(0)a}_{\mu \nu}A^{(m)b\mu}A^{(m)c\nu} \nonumber \\
&&-\frac{1}{4}\left({\cal D}^{(0)ab}_\mu A^{(m)b}_\nu-{\cal D}^{(0)ab}_\nu A^{(m)b}_\mu \right)\left({\cal D}^{(0)ac\mu} A^{(m)c\nu}-{\cal D}^{(0)ac\nu} A^{(m)c\mu} \right)\nonumber \\
&&+\frac{1}{2}\left({\cal D}^{(0)ab}_\mu A^{(m)b}_5\right)\left({\cal D}^{(0)ac\mu} A^{(m)c}_5\right)+\frac{1}{2}m^2_mA^{(m)a}_\mu A^{(m)a\mu}\nonumber \\
&&-\frac{1}{2\xi}\left({\cal D}^{(0)ab}_\mu A^{(m)b\mu} \right)\left({\cal D}^{(0)ac}_\nu A^{(m)c\nu} \right)-\frac{1}{2}\xi m^2_mA^{(m)a}_5A^{(m)a}_5\nonumber \\
&&+\bar{C}^{(m)b}\left({\cal D}^{(0)ab}_\mu {\cal D}^{(0)ac\mu}\right)C^{(m)c}-\xi m^2_m\bar{C}^{(m)a}C^{(m)a}\, .
\end{eqnarray}
In this expression, the first and second terms arise from the $-(1/4){\cal F}^{(0)a}_{\mu \nu}{\cal F}^{(0)a\mu \nu}$ and  $-(1/4){\cal F}^{(m)a}_{\mu \nu}{\cal F}^{(m)a\mu \nu}$ parts of ${\cal L}_{eff}$, respectively. The third and fourth terms come from $-(1/4){\cal F}^{(m)a}_{\mu 5}{\cal F}^{(m)a\mu}_5$.
The fifth and sixth terms are generated by the gauge--fixing part, ${\cal L}_{GF}$. Finally, the seventh and eighth terms are produced by the Faddeev--Popov ghost term, ${\cal L}_{FPG}$. The Feynman rules for the trilinear and quartic vertices $A^{(0)a\alpha}(k_1)A^{(m)b\lambda}(k_2)A^{(n)c\rho}(k_3)$ and $A^{(0)a\alpha} A^{(0)b\beta}A^{(m)c\lambda}A^{(n)d\rho}$ are respectively given by $g\delta^{mn}f^{abc}\Gamma_{\alpha \lambda \rho}(k_1,k_2,k_3)$ and $ig^2\Gamma^{(mn)abcd}_{\alpha \beta \lambda \rho}$, where
\begin{equation}
\Gamma_{\alpha \lambda \rho}(k_1,k_2,k_3)=(k_3-k_2)_\alpha g_{\lambda \rho}-\left(k_1-k_2-\frac{1}{\xi}k_3\right)_\rho g_{\alpha \lambda}+\left(k_1-k_3-\frac{1}{\xi}k_2 \right)_\lambda g_{\alpha \rho}\, ,
\end{equation}
\begin{equation}
\Gamma^{(mn)abcd}_{\alpha \beta \lambda \rho}=\delta^{mn}\left[f^{ade}f^{bce}\left(g_{\alpha \lambda}g_{\beta\rho}-g_{\alpha\beta}g_{\lambda\rho}+\frac{1}{\xi}g_{\alpha\rho}g_{\beta\lambda}\right)+f^{ace}f^{bde}\left(g_{\alpha \rho}g_{\beta\lambda}-g_{\alpha\beta}g_{\lambda\rho}+\frac{1}{\xi}g_{\alpha\lambda}g_{\beta\rho}\right)\right]\, .
\end{equation}
In the above expression, all momenta are pointing to the vertex. Notice that, as a consequence of the invariance under the SGT, the vertex function associated with the trilinear vertex satisfies the following simple Ward identitiy
\begin{equation}
k^\alpha_1 \Gamma_{\alpha \lambda \rho}(k_1,k_2,k_3)=\Gamma^{(m)}_{\lambda \rho}(k_2)-\Gamma^{(m)}_{\lambda \rho}(k_3)\, ,
\end{equation}
where $\Gamma^{(m)}_{\lambda \rho}(k)$ is the two--point vertex function given by:
\begin{equation}
\Gamma^{(m)}_{\lambda \rho}(k)=\left(k^2-m^2_m\right)g_{\lambda \rho}-\left(1-\frac{1}{\xi}\right)k_\lambda k_\rho \,.
\end{equation}

\section{One--loop renormalizability of light Green's functions}
\label{olr}
The structure of the Lagrangian ${\cal L}^{(0)(n)}\left(A^{(0)a}_\mu,A^{(n)a}_\mu,A^{(n)a}_5\right)$ suggests that the only divergences induced by the excited KK modes on light Green's functions at the one--loop level are those already present in the standard Yang--Mills theory and can therefore be absorbed by the parameters of the light theory. We now proceed to show that this is indeed the case. This requires to quantize the standard Yang--Mills theory, which means that a gauge--fixing procedure for the zero mode gauge field $A^{(0)a}_\mu$ must be introduced. The one--loop renormalizability of standard Yang--Mills theories is particulary simple if one uses the Background Field Method~\cite{BFM}, since in this scheme, as already commented, the quantum theory preserves invariance under the SGT. This formal gauge invariance sets powerful constraints on the infinities that can occur in the effective action. We split the zero mode gauge field $A^{(0)a}_\mu$ into a classical background field, $A^{(0)a}_\mu$, and a fluctuating quantum field, ${\cal A}^{(0)a}_\mu$,
\begin{equation}
A^{(0)a}_\mu \to A^{(0)a}_\mu+{\cal A}^{(0)a}_\mu \, .
\end{equation}
The classical part $A^{(0)a}_\mu$ is treated as a fixed field configuration and the fluctuating part ${\cal A}^{(0)a}_\mu$ as the integration variable of the functional integral. The Yang--Mills curvature decomposes as follows:
\begin{equation}
F^{(0)a}_{\mu \nu}\to F^{(0)a}_{\mu \nu}+{\cal D}^{(0)ab}_\mu {\cal A}^{(0)b}_\nu -{\cal D}^{(0)ab}_\nu {\cal A}^{(0)b}_\mu+gf^{abc}{\cal A}^{(0)b}_\mu {\cal A}^{(0)c}_\nu \,.
\end{equation}
As a next step, we choose a gauge--fixing condition that is covariant with respect to SGT of the background gauge field:
\begin{equation}
f^{(0)a}={\cal D}^{(0)ab}_\mu {\cal A}^{(0)b\, \mu}\, .
\end{equation}
Notice that this gauge--fixing procedure is identical to that introduced for the KK gauge modes $A^{(m)a}_\mu$, as both preserve gauge invariance with respect to SGT. The gauge--fixed Lagrangian for the standard Yang--Mills theory is
\begin{eqnarray}
{\cal L}^{(0)}_{YM}&=&-\frac{1}{4}\Big(F^{(0)a}_{\mu \nu}+{\cal D}^{(0)ab}_\mu {\cal A}^{(0)b}_\nu -{\cal D}^{(0)ab}_\nu {\cal A}^{(0)b}_\mu+gf^{abc}{\cal A}^{(0)b}_\mu {\cal A}^{(0)c}_\nu \Big)^2 \nonumber \\
&&-\frac{1}{2\xi}\Big({\cal D}^{(0)ab}_\mu {\cal A}^{(0)b\, \mu}\Big)^2+\bar{C}^{(0)a}\Big({\cal D}^{(0)ab}_\mu {\cal D}^{(0)bd \mu}
+{\cal D}^{(0)ab \mu}f^{bcd}{\cal A}^{(0)c}_\mu \Big)C^{(0)d}\, .
\end{eqnarray}
This Lagrangian is invariant under the SGT, with the fluctuating quantum fields and the ghost fields transforming in the adjoint representation of the group. We now center our attention in the one--loop contribution to light Green's functions of both the zero modes and the excited ones. At this level, only quadratic terms in ${\cal A}^{(0)a}_\mu$ and $A^{(m)a}_\mu$ can contribute. From the ${\cal L}^{(0)}_{YM}$ and ${\cal L}^{(0)(n)}\left(A^{(0)a}_\mu,A^{(n)a}_\mu,A^{(n)a}_5\right)$ Lagrangians, we can see that the one--loop effects of both the zero modes and excited ones are governed by the following Lagrangian:
\begin{equation}
{\cal L}_{1-loop}={\cal L}^{(0)}_{1-loop}+\sum^{\infty}_{m=1}{\cal L}^{(m)}_{1-loop}\, ,
\end{equation}
where
\begin{eqnarray}
{\cal L}^{(0)}_{1-loop}&=&-\frac{1}{2}\left(\frac{1}{2}\left({\cal D}^{(0)ab}_\mu {\cal A}^{(0)b}_\nu -{\cal D}^{(0)ab}_\nu {\cal A}^{(0)b}_\mu\right)^2 +gf^{abc}F^{(0)a\mu \nu}{\cal A}^{(0)b}_\mu {\cal A}^{(0)c}_\nu+\frac{1}{\xi}\left(D^{(0)ab}_\mu {\cal A}^{(0)b\mu}\right)^2\right)\nonumber \\
&& +\bar{C}^{(0)b}\left({\cal D}^{(0)ab}_\mu {\cal D}^{(0)ac\mu}\right)C^{(0)c}\, ,
\end{eqnarray}
\begin{eqnarray}
{\cal L}^{(m)}_{1-loop}&=&-\frac{1}{2}\Bigg(\frac{1}{2}\left({\cal D}^{(0)ab}_\mu A^{(m)b}_\nu -{\cal D}^{(0)ab}_\nu A^{(m)b}_\mu\right)^2 +gf^{abc}F^{(0)a\mu \nu}A^{(m)b}_\mu A^{(m)c}_\nu+\frac{1}{\xi}\left(D^{(0)ab}_\mu A^{(m)b\mu}\right)^2 \nonumber \\
&&-\left(m_mA^{(m)a}_\mu\right)^2\Bigg)+\bar{C}^{(m)b}\left({\cal D}^{(0)ab}_\mu {\cal D}^{(0)ac\mu}-\xi m^2_m\right)C^{(m)c}\nonumber \\
&&+\frac{1}{2}\left(\left({\cal D}^{(0)ab}_\mu A^{(m)b}_5\right)^2+\left(m_mA^{(m)a}_5\right)^2\right)\, .
\end{eqnarray}
Notice the similitude between the ${\cal L}^{(0)}_{1-loop}$ and ${\cal L}^{(m)}_{1-loop}$ terms. In particular, it is important to stress that the couplings appearing in the ${\cal L}^{(m)}_{1-loop}$ Lagrangian are of renormalizable type and are all those that are allowed by gauge invariance. This fact implies that the type of infinities generated by the KK modes $A^{(m)a}_\mu$ must be identical to those generated by the fluctuations associated with the zero mode ${\cal A}^{(0)a}_\mu$. On the other hand, it is a well--known fact that, as a consequence of the gauge invariance associated with the background field gauge, the UV divergence of a pure standard Yang--Mills theory must be of the way
\begin{equation}
{\cal L}^{(0)}_\infty=-\frac{1}{4}L^{(0)}F^{(0)a}_{\mu \nu}F^{(0)a \mu \nu}\, ,
\end{equation}
where it is expected from dimensional analysis that the constant $L^{(0)}$ is logarithmically divergent. Since the complete ${\cal L}_{1-loop}$ Lagrangian is invariant under the SGT, the same type of UV divergence is expected from each KK mode. So, the one--loop UV divergence is of the way
\begin{equation}
{\cal L}_\infty=-\frac{1}{4}L\, F^{(0)a}_{\mu \nu}F^{(0)a \mu \nu}\, ,
\end{equation}
where
\begin{equation}
L=\sum^{\infty}_{m=0}L^{(m)}\, ,
\end{equation}
with $L^{(0)}=L^{(1)}=\cdots L^{(m)}=\cdots $, which, as already mentioned, is a consequence of gauge invariance and also of the fact that the couplings of excited KK modes to the zero ones are identical to those among zero modes only. This in turn implies that no divergences multiplying gauge invariants of canonical dimension higher than four can arise. Then, the UV divergences generated by the KK modes $A^{(m)a}_\mu$ at the one--loop level can be absorbed in the light theory by defining the renormalized fields as follows
\begin{equation}
A^{(0)a\, R}_\mu=\sqrt{1+L}\, A^{(0)a}_\mu \, ,
\end{equation}
which leads to a renormalized curvature given by
\begin{equation}
F^{(0)a\, R}=\partial_\mu A^{(0)a\, R}_\nu-\partial_\nu A^{(0)a\, R}_\mu+g^R f^{abc}A^{(0)b\, R}_\mu A^{(0)c\, R}_\nu\, ,
\end{equation}
where the structure constant is also renormalized by the same factor:
\begin{equation}
g^R=(1+L)^{-1/2}g\, .
\end{equation}
This result arises as a consequence of gauge invariance, which exhibits the particular virtue of the background field gauge and of the covariant gauge conditions that we introduced for the KK excited modes. On the other hand, it is a well known fact from radiative corrections that logarithmically divergent integrals can introduce nondecoupling effects proportional to the logarithm of the mass of the particle circulating in the loop. In our case, associated with the divergent integrals, there would arise terms of the form $log(mR^{-1}/\mu)$, with $\mu$ a mass scale like the one introduced by dimensional regularization. These type of terms do not decouple in the limit of a very small compactification scale, but these effects are unobservable indeed, as they can be absorbed by renormalization. The fact that the UV divergences induced by the excited KK modes at the one--loop level are controllable, opens the possibility of investigating in an unambiguous way the one--loop impact of these excitations on some electroweak observables. In particular, this is important for the case of UED models, as the one studied here, in which the conservation of the discrete momentum $k_5=mR^{-1}$ implies that the KK parity, $(-1)^m$, is conserved and no couplings involving only one single KK mode can arise. This in turn implies that no contributions to the electroweak observables can arise at the tree level~\cite{Apel}. This means that in this class of extra dimensional models, a direct contribution of excited KK modes to low--energy observables first arises at the one--loop level. Although these observables can receive tree level effects from operators of canonical dimension higher than $D(=5)$, some studies carried out on some electroweak observables show that, in theories with only one extra dimension, the one--loop effect dominates. The importance of our result concerning the renormalizability of the one--loop effects of KK modes on light Green's functions can be best appreciated in this context. A very recent calculation for the one--loop form factors of the trilinear electroweak vertices $W^-W^+V$ ($V=\gamma, Z$)~\cite{FMNRT1}, as well as some preliminaries studies for other rare processes, as light by light scattering~\cite{FMNRT2}, indicate that this is indeed the case, as the corresponding amplitudes possess the main properties observed in the context of other renormalizable theories, such as gauge invariance, absence of ultraviolet divergences, and a well behavior at high energies. In particular, the heavy physics effects decouple in the large $R^{-1}$ limit~\cite{FMNRT1,FMNRT2}. This should be compared with the case of nonuniversal extra dimensional (NUED) models, in which some fields are confined to the $4D$ brane. In these class of models, the discrete momentum is not conserved in the brane but only in the bulk~\cite{Bdll}. As a consequence, vertices involving only one KK excited mode can exist and divergences can arise at the tree level, although the involved propagators are finite if only one extra dimension is considered~\cite{Bdll}. However, even in the case of NUED with only one extra dimension, one--loop effects on light Green's functions are cutoff depending~\cite{Bdll}.

As already commented in the introduction, gauge theories in more than four dimensions are nonrenormalizable in the Dyson's sense, so they must be recognized as effective theories that parametrize the low--energy manifestations of a more fundamental theory. Although at the level of the four dimensional theory the coupling constants are dimensionless and the corresponding Lagrangian does not involve interactions of dimension higher than four, the nonrenormalizable character manifest itself through the infinite multiplicity of the KK modes. It is therefore reasonable to expect that two--loop or higher effects of KK modes on light Green's functions cease to be renormalizable, as a new class of couplings among KK modes appearing in the complete ${\cal L}_{4YM}$ Lagrangian arise and, as a consequence, new discrete infinite sums must be considered. However, effective field theories are predictive in a modern sense~\cite{WBooks, Examples,GW}. Although effective theories arising from compactification of extra dimensions incorporate ingredients that are not present in conventional effective formulations of physical theories, such as the chiral approach to strong interactions~\cite{CHASI} or electroweak effective Lagrangians~\cite{GW}, it is worth presenting some comments on this issue. It is not our objective to study renormalizability in a modern sense of general Kaluza--Klein theories, but only to contrast our result concerning the one--loop renormalizability of light Green's functions in a wider context and to explore in a qualitative way the possibility of integrating out the excited KK modes in this special case of UED models with only one extra dimension. Following the spirit of the paper, we will restrict our discussion to a pure Yang--Mills theory in five dimensions, for simplicity. Since the theory is nonrenormalizable in the Dyson's sense, there is no limit for the number of $SU_5(N)$--invariants that can be introduced. So, the five dimensional Lagrangian comprises an infinite series of effective operators:
\begin{equation}
{\cal L}^{eff}_{5YM}=-\frac{1}{4}{\cal F}^a_{MN}(x,y){\cal F}^{a MN}(x,y)+\sum^{\infty}_N \frac{\beta_N g^{N_1}_5}{M^{N_2}}{\cal O}_N({\cal A}^a_M)\, ,
\end{equation}
where the ${\cal O}_N$ are operators of canonical dimension higher than five, $M_s$ is the energy scale at which the new physics first directly manifests itself, and $\beta_{N}$ is a dimensionless parameter that depends on the details of the underlying physics. In the above Lagrangian, it is assumed that all the independent operators that respect the Lorentz and gauge symmetries are included and that each of them appears multiplied by an unknown dimensionless parameter $\beta_i$. The canonical dimension in each term of the series is appropriately  corrected by introducing factors of $g_5$ and $M^{-1}_s$. Operators of higher canonical dimension will be more suppressed because they involve higher powers of the new physics scale $M^{-1}_s$. Once compactified and integrated the fifth dimension, one obtains a four dimensional effective Lagrangian, given by
\begin{equation}
{\cal L}^{eff}_{4YM}={\cal L}_{4YM}\left(A^{(0)a}_\mu,A^{(m)a}_\mu\right)+\sum^\infty_{N>4}\frac{\alpha_N}{M^{N-4}_s}{\cal O}\left(A^{(0)a}_\mu,A^{(m)a}_\mu\right)\, .
\end{equation}
Several comments are in order here. Besides depending on the light degrees of freedom $A^{(0)a}_\mu$, the four dimensional effective Lagrangian depends in addition on the KK degrees of freedom that arise at the compactification scale $R^{-1}$, which is expected to be below from the scale $M_s$ characterizing the more fundamental theory. In this effective Lagrangian, the compactification scale only arises through the masses of the KK modes. It cannot arise as global factors of inverse powers, since this effect is canceled by factors of $g_5$ appropriately introduced in the five dimensional effective Lagrangian. The presence of two different scales in the effective four dimensional Lagrangian, namely the low--energy scale to which are associated the light degrees of freedom (the Fermi scale in the standard model) and the compactification scale that characterizes the KK modes, which would differ substantially in both their own origin and relative values, is a new complication not present in conventional effective field theories. Another interesting aspect of the four dimensional effective Lagrangian is that it is subject to satisfy, besides the usual Lorentz symmetry, both the SGT and the NSGT. According to renormalizability in a modern sense~\cite{WBooks}, one can to carry out radiative corrections (to light observables) using the above effective Lagrangian. New types of infinities can arise, but this does not constitute a serious problem, as the counterterms needed to remove them are already present in the effective Lagrangian. Such divergences simply renormalize the bare coupling constants $\alpha_i$. Indeed, the difficulties encountered in effective theories are not related with the issue of removing infinities, but with the predictability of the theory, by virtue of the presence of a large number of parameters. Despite the fact that the formalism involves, in principle, an infinite number of local operators, only a finite number of them need to be considered in any given calculation; the number of operators which are considered is determined by the required degree of accuracy: for higher precision more terms in the effective lagrangian must be included, and the number of parameters increases.

On the other hand, since in UED models the contributions to low--energy observables first arise at the one--loop level, the interactions depending on excited KK modes in operators of canonical dimension higher than four appearing in ${\cal L}^{eff}_{4YM}$ can be ignored, as their one--loop effects would be quite  suppressed with respect to those induced by the couplings among KK modes appearing in ${\cal L}_{4YM}$. So, the most relevant pieces of the effective Lagrangian are:
\begin{equation}
{\cal L}^{eff}_{4YM}={\cal L}_{4YM}\left(A^{(0)a}_\mu,A^{(m)a}_\mu\right)+\sum^\infty_{N>4}\frac{\alpha_N}{M^{N-4}_s}{\cal O}\left(A^{(0)a}\right)\, ,
\end{equation}
where in the operators of dimension higher than four any dependence on the KK fields $A^{(m)a}_\mu$ has been dropped. However, the dependence on these fields is  maintained in the dimension four ${\cal L}_{4YM}\left(A^{(0)a}_\mu,A^{(m)a}_\mu\right)$ Lagrangian. It is important to notice that the most important effects on a given observable may arise at the tree--level from some effective operators. Depending of the specific spacetime structure of the extra dimensional model, this contribution may be of the same order or even larger than that induced at the one--loop level by the KK modes. It is therefore important that we clarify, as much as possible, the relative importance of the one--loop effects of KK modes on light Green's functions as compared with those induced at the tree--level by operators of higher canonical dimension. As shown in this work and also in some previous studies~\cite{Papa,Apel}, in the context of UED models that involve only one extra dimension, the one-loop contribution of KK modes to light Green's functions is insensitive to the cutoff $M_s$, it depends only on the compactification scale. Since the compactification scale is lower than the fundamental scale, one can expect that in this type of models the one--loop KK contribution strongly competes with that induced at the tree level, as the latter is suppressed by inverse powers of the fundamental scale $M_s$. This means that in this type of models, both types of contributions must be considered. Explicit studies point out in this direction, as calculations~\cite{Apel,FMNRT1} of some electroweak observables show that the one--loop contribution of the KK modes dominates. If the compactification scale $R^{-1}$ is very above of the available energies, it would be desirable to integrate out each heavy field $A^{(m)a}_\mu$ to obtain an effective Lagrangian depending only on the light fields $A^{(0)a}_\mu$:
\begin{equation}
{\cal L}^{eff}_{4YM}\left(A^{(0)a}_\mu \right)=-\frac{1}{4}F^{(0)a}_{\mu \nu}F^{(0)a\mu \nu}+\sum^{\infty}_{m=1}\sum^\infty_{N>4}\frac{\hat{\alpha}_N}{\left(\frac{m}{R}\right)^{N-4}}\hat{{\cal O}}_N\left(A^{(0)a}\right)+\sum^\infty_{N>4}\frac{\alpha_N}{M^{N-4}_s}{\cal O}\left(A^{(0)a}\right)\, .
\end{equation}
The one--loop renormalizability of the light Green's functions showed above for this Yang--Mills theory with only one extra dimension, suggests that the derivation of this effective Lagrangian is feasible~\cite{NT}. However, it is not clear if this is possible for more general formulations of theories in extra dimensions, as UED models with more than one extra dimension or NUED with one or more extra dimensions.

\section{Summary}
\label{C}We have studied the gauge structure and quantization of a gauge system that arises after compactification of a pure Yang--Mills theory in five dimensions, with the fifth dimension compactified on the orbifold $S^1/Z_2$ of radius $R$. The importance of studying the role played by the gauge parameters of the compactified theory was stressed through the paper, as they are essential pieces within the context of the BRST symmetry, both at the classical and quantum levels. In our opinion, this issue, which is fundamental to quantize the theory, has not been properly studied. Depending on whether the gauge parameters propagate or not in the fifth dimension, two scenarios can arise. The scenario with the gauge parameters propagating in the bulk leads to a four dimensional theory with a complicated gauge structure due to the presence of an infinite tower of gauge parameters that arise after compactification. In this scenario, we have derived a four dimensional Lagrangian, ${\cal L}_{4YM}$, that differs substantially from the one known in the literature. This is one of our main results. We showed that this Lagrangian satisfies separately two types of gauge transformations, which we called SGT and NSGT. Under the SGT, which are defined by the zero modes of the gauge parameters, the zero modes of the gauge fields transform as gauge fields, whereas the excited KK gauge modes and the pseudo Goldstone bosons transform as matter fields in the adjoint representation of the group. On the other hand, under the NSGT, which are defined by the infinite tower of excited modes of the gauge parameters, the zero modes of the gauge fields are mapped into excited modes in a way that resembles the ordinary adjoint representation, whereas the excited KK gauge modes transform as gauge fields through a gauge--structure tensor similar to that of the SGT, which however involves the covariant derivative of the SGT instead of the ordinary derivative mixing then the zero modes with the excited ones. As far as the pseudo Goldstone bosons are concerned, they transform without mixing with other fields, but in a complicated way. It looks like a combination of a translation plus a rotation. Related to such unphysical fields, we have found a particular gauge transformation which allows us to remove them of the theory. The NSGT are characterized by involving an infinite sum of the mentioned terms. Special emphasis was put on the gauge--structure tensor characterizing the NSGT of the excited KK gauge modes, as its precise determination is fundamental to quantize these gauge fields. It was shown that in order to obtain a four dimensional Lagrangian that respects simultaneously both the SGT and the NSGT, the curvatures must be considered as the fundamental objects in the sense of expanding them in Fourier series instead of the gauge fields, which has been the route followed in the literature. The Lagrangian so obtained is simpler than the one given in the literature, as it is made of contractions among covariant objects (curvatures) that transform in a well--defined way under both the SGT and the NSGT. The gauge--structure tensors associated with the NSGT were derived via three different ways: from the five dimensional transformation laws, by employing the Dirac's method together with the Castellani's gauge generator, and indirectly by using the master equation. It was shown that the theory is subject to first class constraints, showing that the zero modes as well as the excited ones are gauge fields. These constraints, which are not known in the literature, were derived in two ways; applying the Dirac's method to the four dimensional theory and by compactification of the corresponding constraints at the five dimensional level, finding a perfect agreement. As far as the quantization of the theory was concerned, we focused on the quantization of the KK gauge modes, as it is interesting to investigate the loop effects of excited KK modes on light Green's functions. A proper solution of the master equation in five dimensions was used to derive the counterpart of the four dimensional theory. This solution was used to derive the quantum effective action, showing explicitly the structures of the gauge--fixing and Faddeev--Popov ghost terms. A gauge--fixing procedure that is covariant under the SGT was introduced. This covariant quantization scheme, which can greatly simplify the radiative corrections to light Green's functions, is other of our main results. The Lagrangian linking up the light physics with the heavy physics, ${\cal L}^{(0)(n)}\left(A^{(0)a}_\mu,A^{(n)a}_\mu,A^{(n)a}_5\right)$, was presented. This Lagrangian, which is made of five pieces that are separately invariant under the SGT, has all the ingredients of a predictive theory, as it has all the dimension--four interactions that are compatible with the SGT. In addition, it was possible to endow the quantum theory with a $R_\xi$--gauge that is covariant under the SGT, which in practical loop calculations permits a better control of divergencies than in conventional gauges. The gauge structure of this Lagrangian suggests that the only divergences induced by the excited KK modes on light Green's functions would be those already present in the standard Yang--Mills theory and can therefore be absorbed by the parameters of the light theory. A notable attribute of the gauge-fixing Lagrangian presented here is that the fixation of the gauge can be implemented for the KK excitations while keeping the invariance under the SGT, with the possibility of removing such an invariance by employing the usual fixation of the gauge for the standard Yang-Mills theory defined in four dimensions. The existence of two sorts of parameters defining different gauge transformations embodies a symptom of the indepen- dence of these two gauge-fixing procedures with respect to each other.

We took advantage of the possibility of defining independent gauge--fixing procedures for the two types of gauge symmetries, SGT and NSGT, to quantize the standard Yang--Mills theory via the introduction of the background field gauge, which preserves gauge invariance with respect to the SGT. The quantization of the complete theory in a scheme that is covariant under the SGT was necessary in order to show that the KK effects on the light Green's functions are renormalizable at the one--loop level. It was shown that this gauge invariance sets powerful constraints on the infinities that can occur at the one--loop level. In particular, it was shown that only one type of infinities can be generated and that it is the same for the zero modes $A^{(0)a}_\mu$ and the excited ones $A^{(m)a}_\mu$. Due to this, such divergences can be absorbed in a renormalization of the zero mode field $A^{(0)a}_\mu$, showing that the one--loop effects of KK modes on light Green's functions are insensitive to the cutoff $M_s$. The relative importance of the one--loop effects of KK modes on light Green's functions in the context of UED models with only one extra dimension was stressed. In this type of models, this contribution is more important than the one that could be induced at the tree--level by operators of higher canonical dimension and is the first direct contribution of the KK modes to low--energy observables, as they cannot contribute at the tree--level.

The other studied scenario arises from assuming that the gauge parameters do not propagate in the fifth dimension. Since in this scenario both the five dimensional theory and the four dimensional one are governed by the same gauge group, $SU(N)$, no other type of gauge transformations than the SGT can exist. In this case, it makes sense to integrate the fifth dimension by expanding in the action the gauge fields instead of the curvatures, as in the previous scenario. Since in the context of the BRST symmetry the gauge parameters do not appear in the original action $S_0$, but at the level of the extended one $S$, no role is played by the gauge parameters when the fifth dimension is integrated. Only the objects (gauge fields or curvatures) considered as fundamental in the Fourier series determine the result. Due to this, we expected to reproduce the results already known in the literature, but our four dimensional Lagrangian, $\hat{\cal L}_{4YM}$, present some differences. Nonetheless, we found that both the ${\cal L}_{4YM}$ and the $\hat{{\cal L}}_{4YM}$ Lagrangians, as well as the results given in the literature, contains the original part of the ${\cal L}^{(0)(n)}\left(A^{(0)a}_\mu,A^{(n)a}_\mu,A^{(n)a}_5\right)$ Lagrangian, \textit{i.e.} the term that results of removing from ${\cal L}^{(0)(n)}\left(A^{(0)a}_\mu,A^{(n)a}_\mu,A^{(n)a}_5\right)$ the gauge--fixing and Faddeev--Popov terms. In other words, the part of ${\cal L}^{(0)(n)}\left(A^{(0)a}_\mu,A^{(n)a}_\mu,A^{(n)a}_5\right)$ present in the original action $S_0$ is already known in the literature and arises in the two scenarios considered here. However, it is important to stress that in the scenario from which arises ${\cal L}_{4YM}$, the excited KK modes $A^{(m)a}_\mu$ are gauge fields, so diverse propagators can be used in radiative corrections by choosing a particular gauge. Nevertheless, in the scenario from which arises $\hat{{\cal L}}_{4YM}$, the excited KK modes $A^{(m)a}_\mu$ are not gauge fields but matter or Proca fields, so radiative corrections must be calculated in this case using the unitary propagator. As it was emphasized previously, this scenario is unattractive due to presence of massless scalar fields, which in this case cannot be removed of the theory, as they are not pseudo Goldstone bosons. It was shown that this system is subject to both first class and second class constraints, as must be.

In conclusion, in this paper, a quantization procedure for the excited KK gauge modes of a compactified Yang--Mills theory that is covariant under the standard gauge transformations of $SU(N)$ was presented. The effective quantum Lagrangian that links the heavy physics with the light physics, which is invariant under the $SU(N)$ group, was presented. The gauge structure of this Lagrangian suggests that the only divergences generated by the excited KK modes on light Green's functions are those that can be absorbed by the parameters of the light theory. A gauge covariant quantization of the complete theory was used to show that the light Green's functions are renormalizable at the one--loop level. We stress that this is an important result, as it would be possible to predict in an unambiguous way the one--loop radiative corrections of extra dimensions on some electroweak observables, which will be the subject of experimental scrutiny at the next generation of linear colliders.

\acknowledgments{ We acknowledge financial support from CONACYT and SNI (M\' exico).}

\end{document}